\pdfoutput=1
\documentclass[reprint,aps,pre,twocolumn]{revtex4}
\usepackage{amsmath,amsfonts,bm, color,amssymb}
\usepackage{graphicx,epsfig,overpic,hyperref,ulem}
\usepackage{comment}

\usepackage{mathtools}
\usepackage{empheq}
\usepackage{multirow}
\usepackage{tabulary}
\usepackage{caption}
\usepackage{subfig}

\DeclareMathAlphabet\mathbfcal{OMS}{cmsy}{b}{n}
\allowdisplaybreaks

\newcommand{\refeqq}[1]{Eq. \ref{#1}}

\input epsf

\newcommand{\bE}{{\bf E}}
\newcommand{\bP}{{\bf{P}}}
\newcommand{\bQ}{{\bf{Q}}}
\newcommand{\bU}{{\bf{U}}}
\newcommand{\bG}{{\bf{G}}}

\newcommand{\bO}{{\mathbfcal O}}

\newcommand{\bT}{{\bf T}}

\newcommand{\bI}{{\bf I}}

\newcommand{\bOmg}{{\mathbf{\Omega}}}
\newcommand{\bR}{{\bf R}}
\newcommand{\bF}{{\bf F}}

\newcommand{\tvdots}{[\cdot]^3}

\newcommand{\tbr}{\tilde{\br}}

\newcommand{\tbR}{\widetilde{\bR}_{ij}}
\newcommand{\tbF}{\widetilde{\bF}}
\newcommand{\tbE}{\widetilde{\bE}}

\newcommand{\tbP}{\widetilde{\bP}}
\newcommand{\tbQ}{\widetilde{\bQ}}

\newcommand{\tbOmg}{{\widetilde{\bf{\Omega}}}}

\newcommand{\tbT}{\widetilde{\bT}}

\newcommand{\hbR}{\widehat{\bR}_{ij}}
\newcommand{\bbR}{\overline{\bR}_{ij}}

\newcommand{\bzero}{{\bf 0}}

\newcommand*{\tran}{^{\mkern-1.5mu\mathsf{T}}}

\newcommand{\bJ}{{\bf J}}

\newcommand{\bn}{{\bf n}}

\newcommand{\bu}{{\bf u}}
\newcommand{\br}{{\bf r}}

\begin{document}

\title{Colloidal particle electrorotation in a  non-uniform electric field}
\author{ Yi Hu, Petia M. Vlahovska, Michael J. Miksis}
 \email{miksis@northwestern.edu}
\affiliation{%
 Engineering Science and Applied Mathematics, Northwestern University, Evanston, IL 60208
}

\date{\today}

\begin{abstract}
A model to study the dynamics of  colloidal particles in nonuniform electric fields is proposed. For an isolated sphere, the conditions and  threshold for sustained (Quincke) rotation in a linear direct current (dc) field are determined. Particle dynamics becomes more complex with increasing electric field strength, changing from steady spinning around the particle  center to time-dependent orbiting motion around the minimum field location.  Pairs of particles exhibit intricate trajectories,  which are combination of translation, due to dielectrophoresis, and rotation, due to the Quincke effect. Our model provides a basis to study the collective  dynamics of many particles in a general  electric field. 


\end{abstract}

\maketitle

\section{Introduction}
The spontaneous rotation of a particle in a uniform electric field, first observed over a century  ago \cite{Quincke:1896} and now known as Quincke rotation, has been subject of increasing interest in recent years.
An isolated sphere  \cite{Cebers:2000, Lemaire:2002}  or ellipsoid \cite{Dolinsky-Elperin:2009, Quentin:2017a} displays various rotational motions including chaotic reversal of the direction of rotation. Even more complex dynamics is found in a collection of particles. A pair of spheres can undergo intricate trajectories \cite{Das-Saintillan:2013, Dolinsky-Elperin:2012,Lushi-Vlahovska:2014},  large populations can self-organize  in dynamic  patterns \cite{Bartolo:2013, Bartolo:2015, Belovs:2014, Yeo-Lushi-Vlahovska:2014, Yeo-Lushi-PV:2016, Kokot:2015, Snezhko:2016}, and a suspension can exhibit lower effective viscosity \cite{Cebers:2004, Lemaire:2008, Huang-Zahn-Lemaire:2011}  or increased conductivity \cite{Lemaire:2009b} compared to the suspending fluid.

While the  Quincke rotation of an isolated particle in a uniform electric field is well understood \cite{Jones:1984,Turcu:1987, Lemaire:2002}, the collective dynamics of many Quincke rotors is a largely unexplored problem.
Its modeling is particularly challenging because the induced dipole of a particle is affected by the presence of other particles. Hence, the question arises - how is   electrorotation  affected by nonuniformities in the electric field (either due to the presence of other particles, or applied by design, i.e., using complex arrangement of electrodes)? A non-uniform field also induces dielectrophoresis  \cite{Pohl:1951, Jones:1996, Jones:2003}, hence the overall particle dynamics becomes a complex mix of translation and rotation. 

In this paper we explore the dynamics of spheres in a nonuniform DC electric field. In the case of an isolated sphere, we derive the threshold for Quincke rotation in a  linear field.  In the case of a pair of spheres, we identify the evolution equations for the multipolar moments  (dipole and quadrupole) and particle positions. 
The model can be generalized to many particles and arbitrary non-uniform fields.  
 We present numerical results illustrating interesting particle dynamics for single, pair and multi-particle configurations. 
 
\section{Problem formulation}

\subsection{Electrostatic field}
Consider an isolated spherical particle
with permittivity $\epsilon_p$ and conductivity $\sigma_p$
 suspended in a homogeneous fluid with  permittivity $\epsilon_f$ and conductivity $\sigma_f$.  We adopt the leaky dielectric model \cite{Melcher-Taylor:1969}, which assumes a charge-free bulk. Accordingly,  the electric potentials satisfy the Laplace equation, i.e, $\nabla^2 \phi=0$, with the electric field defined as $\bE=-\nabla\phi$. If  the  applied electric field  is $\bE_\infty=-\nabla\phi_a$, 
the total electric potential can be written as $\phi=\phi_a+\phi_d$ and  the perturbation in the electric field due to the presence of the sphere  $\phi_d$ can be written as a multipolar expansion in $r$ as,
\begin{equation}
\label{pt3}
\begin{split}
	\phi_d(\mathbf{r})&=\frac{\mathbf{r\cdot P}}{r^3}+\frac{1}{2}\frac{\mathbf{rr: Q}}{r^5}+\cdots, \qquad r>a,\\
	\bar \phi_d(\mathbf{r})&=\frac{\mathbf{r\cdot P}}{a^3}+\frac{1}{2}\frac{\mathbf{rr: Q}}{a^5}+\cdots, \qquad r\leq a.
\end{split}	
\end{equation} 
where $r=|\mathbf{r}|$, $\bP$ and $\mathbf{Q}$ are the dipole and quadrupole moments, and $a$ is the sphere radius. The coordinates system is centered at the sphere. 

Ohmic currents from the bulk, $\bJ=\sigma \bE$, charge the interface and give rise to induced free charge
 $q=\mathbf{n}\cdot [\epsilon_f\bE-\epsilon_p\bar \bE]$. Here $\bar \bE=-\nabla(\phi_a+\bar\phi_d)$ is evaluated on the particle side of the interface.
In addition to conduction, the  induced charge is affected by  convection due to the particle rotation
\begin{equation}
\label{qc}
	\frac{\partial q}{\partial t}+\mathbf{n}\cdot[\bJ-\bar \bJ]+\nabla_s \cdot(q\mathbf{u}_s)=0 \quad \text{at} \quad r=a,
\end{equation}
where $\mathbf{n}$ is the unit normal vector, $\nabla_s=(\mathbf{I-nn})\cdot \nabla$ and $\mathbf{u}_s$ is the velocity of a point on the particle surface. In a frame of reference translating with the particle, the surface motion is pure rotation $\Omega$ and hence $\mathbf{u}_s=\mathbf{\Omega}\times a \mathbf{n}$.

If the applied field spatial variation  on the particle scale is small,
then $\phi_a$ can be linearized around the particle center
\begin{equation}
\label{pt2}
	\phi_a(\mathbf{r})=\phi_a(\bzero)+\br\cdot \nabla\phi_a(\bzero)+\cdots
	\end{equation} 
Combining the multipole expansion for the electric potential, \refeqq{pt3}, and  \refeqq{pt2}, in the charge conservation equation \refeqq{qc}  yields the evolution equations for the particle dipole and quadruple moments
\begin{equation}
\label{p1}
\frac{d \bP}{dt}=\mathbf{\Omega\times [P}+a^3\epsilon_{cm}\nabla\phi_{a}(0)]-\frac{1}{\tau_{mw}}[\bP+a^3\sigma_{cm}\nabla\phi_{a}(0)],
\end{equation}
where
\begin{equation}
\epsilon_{cm}=\frac{\epsilon_p-\epsilon_f}{\epsilon_p+2\epsilon_f}, \quad \sigma_{cm}=\frac{\sigma_p-\sigma_f}{\sigma_p+2\sigma_f},\quad \tau_{mw}=\frac{\epsilon_p+2\epsilon_f}{\sigma_p+2\sigma_f},\nonumber
\end{equation}
and
\begin{equation}
\label{qc1}
\begin{split}
\frac{d \mathbf{Q}}{dt}=&\mathbf{\{\Omega \times  [Q}+2a^5\epsilon'_{cm}\nabla\nabla\phi_a(0)]\}\\
&+\mathbf{\{\Omega \times  [Q}+2a^5\epsilon'_{cm}\nabla\nabla\phi_a(0)]\}\tran\\
&-\frac{1}{\tau'_{mw}}\mathbf{[Q}+2a^5\sigma'_{cm}\nabla\nabla\phi_a(0)].
\end{split}
\end{equation}
where
\begin{equation}
\epsilon'_{cm}=\frac{\epsilon_p-\epsilon_f}{2\epsilon_p+3\epsilon_f}, \quad \sigma'_{cm}=\frac{\sigma_p-\sigma_f}{2\sigma_p+3\sigma_f}, \quad \tau'_{mw}=\frac{2\epsilon_p+3\epsilon_f}{2\sigma_p+3\sigma_f},\nonumber
\end{equation}
and superscript $T$ denotes  transpose. 
Details of the derivation of \refeqq{p1} and \refeqq{qc1} can be found in Appendix A. Note that even though $\mathbf{\Omega \times  Q}$ doesn't have to be symmetric at all time, given a traceless and symmetric $\mathbf{Q}$ initially, \refeqq{qc1} preserves symmetry and zero trace. In the absence of rotation,  the dipole and quadrupole moments  relax toward the steady state with two slightly different Maxwell-Wagner times $\tau_{mw}$ and $\tau'_{mw}$, which depend on material electric properties. 
The evolution of higher order moments in the expansion \refeqq{pt3} can be obtained in a similar way.  However, in a linear applied electric field (i.e.,  a spatially slowly varying external electric field) these contributions come at a higher order and are negligible in the far-field approximation.

The force and torque on the particle are calculated by the effective multipole moment method \cite{Jones:1996, Jones:2003}, which at the order of our approximation gives
\begin{equation}
\label{ft1}
\begin{split}
	\mathbf{F}^{el}&=-4\pi\epsilon_f (\bP\cdot \nabla \nabla\phi_a(\bzero)+\frac{1}{6}\mathbf{Q}:\nabla\nabla \nabla\phi_a(\bzero)),\\
	\mathbf{T}^{el}&=-4\pi\epsilon_f (\bP\times \nabla\phi_a(\bzero)+(\mathbf{Q}\cdot \nabla)\times \nabla\phi_a(\bzero)).
\end{split}
\end{equation}
Note that the above expressions are strictly valid for an isolated sphere in a linear applied field.

\subsection{Particle motion}

For small  particles, typically  inertia is negligible. Accordingly, the translational  velocity, $\bU$,  and rotational rate, $\bOmg$, of a sphere  is determined by the balance of  electrostatic force and Stokes drag,
\begin{equation}
\label{mt1}
\begin{split}
	\mathbf{F}^{el}&=6\pi\eta_f a (\mathbf{-u^{\infty}-\nabla^2\bu^\infty+\bU}),\\
	\mathbf{T}^{el}&=8\pi\eta_f a^3 (\mathbf{-\Omega^{\infty}+\Omega}),
\end{split}
\end{equation}
where $\eta_f$ is the viscosity of the suspending fluid, and $\mathbf{u^{\infty}}$ (and $\mathbf{-\Omega^{\infty}}$)  is a background flow (either applied or generated by the motion of other spheres, if present) evaluated at the sphere center.
In our study, the background flow is zero for an isolated particle. In the case of multiple particles,  the background flow is the flow induced by the motion of the rest of the particles.

\section{An isolated sphere  in a linear electric field}
\subsection{Threshold for electrorotation}
The classic Quincke  electrorotation considers an isolated  sphere suspended in a homogeneous fluid and exposed to uniform DC electric  field $\mathbf{E}_\infty$. The threshold field for electrorotation is given by \cite{Jones:1984, Turcu:1987},
\begin{equation}
\label{ec}
|\mathbf{E}|>E_c=\sqrt{\frac{2\eta_f}{\epsilon_f\tau_{mw}(\epsilon_{cm}-\sigma_{cm})}},
\end{equation}
and the  rotation rate is
\begin{equation}
\label{uniomg}
|\mathbf{\Omega}|=\pm\frac{1}{\tau_{mw}}\sqrt{\left(\frac{|\bE_{\infty}|}{E_c}\right)^2-1}.
\end{equation}
 \refeqq{ec} shows that electro-rotation can occur only if  $\epsilon_{cm}>\sigma_{cm}$.

In an non-uniform field, this criterion can be generalized. \refeqq{p1} and \refeqq{qc1}
show that the steady dipole and symmetric quadrupole moments satisfy,
\begin{equation}
\label{omgs1}
\begin{split}
	&\mathbf{\Omega} \times [\bP+a^3\epsilon_{cm}\nabla \phi_a]-\frac{1}{\tau_{mw}}[\bP+a^3\sigma_{cm}\nabla \phi_a]=\bzero\\
	&\mathbf{\Omega} \times [\mathbf{Q}+2a^5\epsilon'_{cm}\nabla\nabla \phi_a]-\frac{1}{2\tau'_{mw}}[\mathbf{Q}+2a^5\sigma'_{cm}\nabla\nabla \phi_a]=\bzero.
\end{split}
\end{equation}
By taking inner and outer product of \refeqq{omgs1} with $\mathbf{\Omega}$, we  obtain $\bP$ and $\bQ$ in terms of $\bOmg$,
\begin{equation}
\label{pqs1}
\begin{split}
	\bP&=A_1[\mathbf{\Omega}\times \nabla \phi_a+\tau_{mw}(\mathbf{\Omega}\cdot \nabla \phi_a)\mathbf{\Omega}]-A_2\nabla \phi_a\\
	\mathbf{Q}&=A_3[\mathbf{\Omega}\times \nabla\nabla \phi_a+2\tau'_{mw}(\mathbf{\Omega}\cdot \nabla\nabla \phi_a)\mathbf{\Omega}]-A_4\nabla \nabla\phi_a,
\end{split}
\end{equation}
where the coefficients are,
\begin{equation}
\begin{split}
&A_1=\frac{a^3\tau_{mw}(\epsilon_{cm}-\sigma_{cm})}{1+\Omega^2\tau^2_{mw}}, A_2=a^3[\epsilon_{cm}-\frac{\epsilon_{cm}-\sigma_{cm}}{1+\Omega^2\tau^2_{mw}}], \\
&A_3=\frac{2a^5\tau'_{mw}(\epsilon'_{cm}-\sigma'_{cm})}{1+4\Omega^2\tau'^2_{mw}}, A_4=2a^5[\epsilon'_{cm}-\frac{\epsilon'_{cm}-\sigma'_{cm}}{1+4\Omega^2\tau'^2_{mw}}].
\end{split}
\end{equation}
Substituting \refeqq{pqs1} back into the torque balance equation \refeqq{mt1}, and taking the inner product with $\bOmg$, we find an equation for $\bOmg$ 
\begin{equation}
\label{quincke}
\begin{split}
	2\eta_f |\mathbf{\Omega}|^2&=\epsilon_f\frac{\tau_{mw}(\epsilon_{cm}-\sigma_{cm})}{1+|\mathbf{\Omega}|^2\tau_{mw}^2}[-(\mathbf{\Omega}\cdot\bE_{\infty})^2+|\mathbf{\Omega}|^2|\bE_{\infty}|^2]\\
	& +\epsilon_f\frac{4\tau'_{mw}a^2(\epsilon'_{cm}-\sigma'_{cm})}{1+4|\mathbf{\Omega}|^2\tau'^2_{mw}}[-|\mathbf{\Omega}\cdot\nabla \bE_{\infty}|^2\nonumber\\
	&+|\mathbf{\Omega}|^2\|\nabla \bE_{\infty}\|^2].
\end{split}
\end{equation}
Here we denote $\|\mathbf{T}\|^2=\sum T^2_{ij}$. Unlike  the uniform field case, in a non-uniform field  $\mathbf{\Omega}\cdot\bE_{\infty}$ in general is nonzero.
 \refeqq{quincke} yields a trivial solution,
\begin{equation}
	\mathbf{\Omega}=0.
\end{equation}
A nontrivial solution of \refeqq{quincke} requires that
\begin{equation}
\label{quincke2}
\begin{split}
	2\eta_f &=\epsilon_f\frac{\tau_{mw}(\epsilon_{cm}-\sigma_{cm})}{1+|\mathbf{\Omega}|^2\tau_{mw}^2}(|\bE_{\infty}|^2-\frac{|\mathbf{\Omega}\cdot\bE_{\infty}|^2}{|\mathbf{\Omega}|^2})\\
	& +\epsilon_f\frac{4\tau'_{mw}a^2(\epsilon'_{cm}-\sigma'_{cm})}{1+4|\mathbf{\Omega}|^2\tau'^2_{mw}}(\|\nabla\bE_{\infty}\|^2-\frac{|\mathbf{\Omega}\cdot\nabla\bE_{\infty}|^2}{|\mathbf{\Omega}|^2}).
\end{split}
\end{equation}
Although this equation cannot be solved  explicitly for $\Omega$, we can identify a necessary condition for the existence of a nontrivial $\Omega$. In particular, when $\epsilon_{cm}-\sigma_{cm}$ and $\epsilon'_{cm}-\sigma'_{cm}$ are both non-negative, we obtain that
\begin{align}
\label{quincke3}
	2\eta_f &\leq \epsilon_f\tau_{mw}(\epsilon_{cm}-\sigma_{cm})|\bE_{\infty}|^2\nonumber\\
	&+4\epsilon_f\tau'_{mw}a^2(\epsilon'_{cm}-\sigma'_{cm})\|\nabla\bE_{\infty}\|^2.
\end{align}
should be satisfied for  electrorotation to occur.

As an example of a nonuniform electric field  let us consider 
\begin{equation}
\label{eap}
	\mathbf{E}_{\infty}=\alpha(x\mathbf{i}-z\mathbf{k}),
\end{equation}
which  is experimentally generated by a hyperbolic cylinder electrodes; here $\mathbf{i}$ and $\mathbf{k}$ are the unit vectors in the $x$ and $z$ direction. The parameter $\alpha$ ia a measure of the field gradient. It should be noted that if a constant field were added to \refeqq{eap} for nonzero $\alpha$, the effect would simply be a translation of the coordinate system, and hence this term is not included here.

From \refeqq{omgs1}, \refeqq{pqs1}, \refeqq{quincke2} and \refeqq{eap}, we obtain for a particle centered at $x=0$ and $z=0$, a steady rotation around the $y$ axis, with the magnitude
\begin{equation}
\label{omgL0}
\begin{split}
|\bOmg|=\sqrt{\frac{\epsilon_f(\epsilon'_{cm}-\sigma'_{cm})a^2\alpha^2}{\eta_f\tau'_{mw}}-\frac{1}{4\tau'^2_{mw}}}.
\end{split}
\end{equation}

Introduce $G=a\alpha/E_c$, this is a measure of the field strength over the particle. With this definition, we can identify a critical value of $G$, $G^*$, in \refeqq{omgL0} which generates a nonzero rotation predicted when $|\bOmg|=0$, 
\begin{equation}
\label{ls3}
\begin{split}
	G^*=\frac{1}{4}\sqrt{\frac{\tau_{mw}(\epsilon_{cm}-\sigma_{cm})}{\tau'_{mw}(\epsilon'_{cm}-\sigma'_{cm})}}.
\end{split}
\end{equation}
This is the threshold for electorotation in a linear field in the whole plane. As shown later, $G^*\approx 0.4092$ for the physical system considered here.

\subsection{Dynamics and particle trajectories}

Here we consider particle motion in the nonuniform electric field \refeqq{eap}. Since the applied field direction is parallel to the $x-z$ plane, particle motion is expected to be confined to  the $x-z$ plane. Still, this special case will allow us to illustrate the effect of non-uniformities on the particle dynamics.  Note that although the evolution equations \refeqq{p1}, \refeqq{qc1} and \refeqq{mt1} are derived for a slowly varying electric field, they are exact for the special case \refeqq{eap}.

Henceforth we non-dimensionalize all variables by the drop radius $a$, electrohydrodynamic time $t_{ehd}={\eta_f}/{(\epsilon_f E^2_c)}$ and electric field threshold for Quincke rotation $E_c$ given by \refeqq{ec}. All dimensionless variables are denoted by ``tilde''.
The dimensionless  \refeqq{p1} and \refeqq{qc1} are
\begin{equation}
\label{nd1}
\begin{split}
\frac{d \tbP}{d\tilde{t}}=&\tbOmg \times [\tbP-\epsilon_{cm}G(\hat{x}\mathbf{i}-\hat{z}\mathbf{k})]\\
&-\frac{1}{D}[\tbP-\sigma_{cm}G(\hat{x}\mathbf{i}-\hat{z}\mathbf{k})],\\
\end{split}
\end{equation}
\begin{equation}
\label{nd2}
\begin{split}
\frac{d \tbQ}{d\tilde{t}}=&\tbOmg\times [\tbQ-2\epsilon'_{cm}\widetilde{\mathcal{G}}]\\
&+[\tbOmg\times [\tbQ-2\epsilon'_{cm}\widetilde{\mathcal{G}}]]\tran\\
&-\frac{1}{D'}[\tbQ-2\sigma'_{cm}\widetilde{\mathcal{G}}],
\end{split}
\end{equation}
where 
\begin{equation}
D=\frac{\tau_{mw}}{t_{ehd}}=\frac{\epsilon_{f}E^2_c\tau_{mw}}{\eta_f}, D'=\frac{\tau'_{mw}}{t_{ehd}}=D\frac{\tau'_{mw}}{\tau_{mw}},\nonumber
\end{equation}
and
\begin{equation}
\widetilde{\mathcal{G}}=G\begin{pmatrix} 
1 & 0 &0\\
0&0&0\\
0&0 & -1 
\end{pmatrix}.
\end{equation}
Note that the applied field in \refeqq{p1} and \refeqq{qc1} are evaluated at the particle center ($\hat x=\hat y=0$), but for a moving particle in a nonuniform field it should be evaluated at the current particle location. Hence there is a spacial dependence in the electric field in \refeqq{nd1}.

The dimensionless \refeqq{mt1}  yields for the particle translation and rotation
\begin{equation}
\label{gn2}
\begin{split}
	\tilde{\mathbf{U}}&=\frac{2 }{3}\tbP\cdot \widetilde{\mathcal{G}},\\
	\widetilde{\bOmg} &=\frac{1}{2}[\tbP\times \frac{E_0}{E_c}(\tilde{x}\mathbf{i}-\tilde{z}\mathbf{k})+\tbQ\times\widetilde{\mathcal{G}}].
\end{split}
\end{equation}

For the numerical calculations,  we chose the experimental system of  Ref. \cite{Lemaire:2008}: $\epsilon_{mw}=-0.1092$, $\epsilon'_{mw}=-0.0670$, $\sigma_{mw}=-0.5$ and $\sigma'_{mw}=-0.3333$, $E_c=827.3V/mm$, $\tau_{mw}=2.94$ms, $\tau'_{mw}=3.20$ms, $D=5.1520$, $D'=5.6054$. From \refeqq{ls3} we find that to guarantee nonzero electrorotation in the whole plane we need $G^*>0.4092$.

\begin{figure}[h!]
{\centerline{\includegraphics[width=2.8in,height=4.5in]{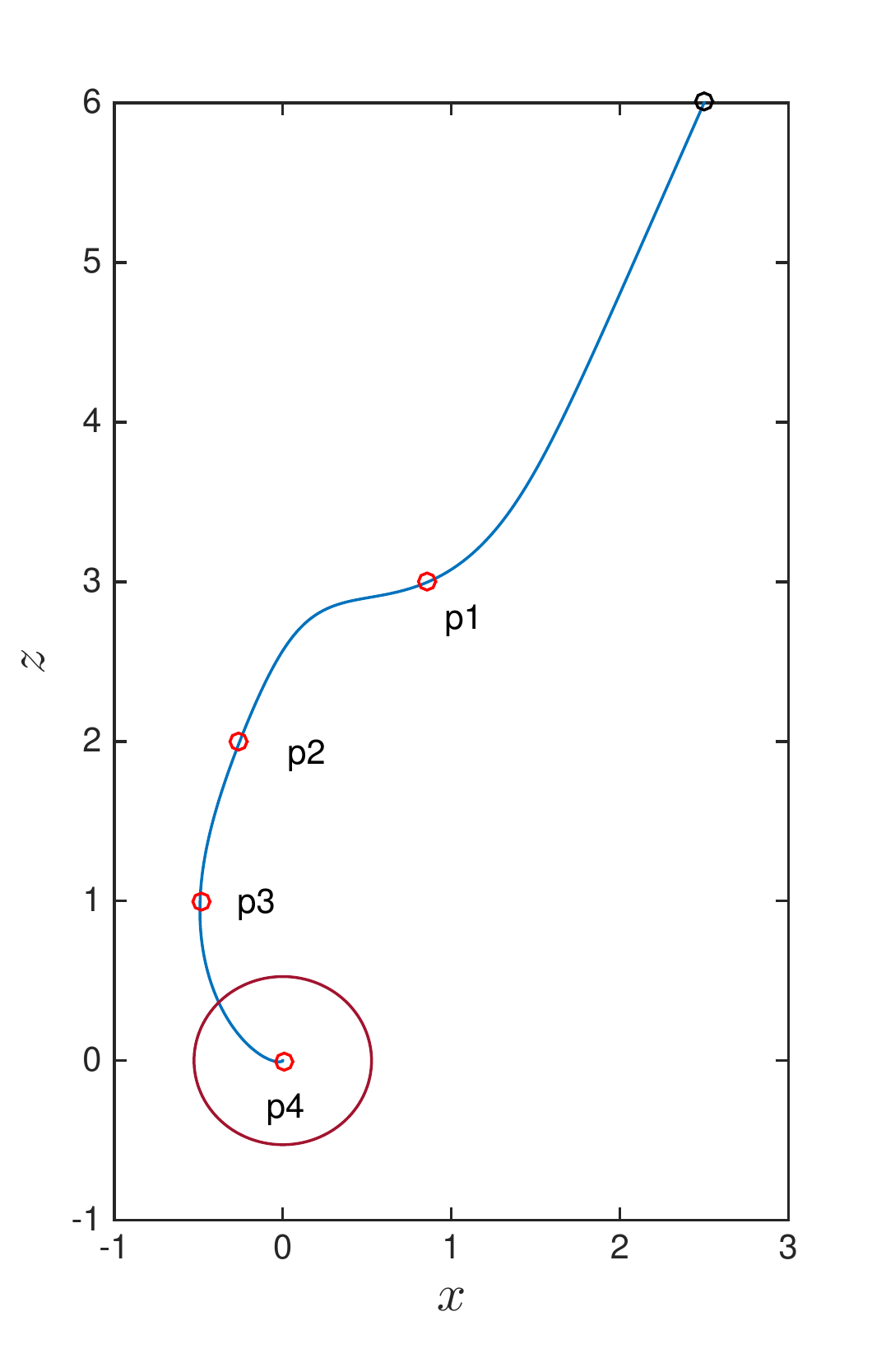}}}
\caption{\footnotesize  One particle trajectory starting at $x=2.5$, $z=6.0$. Initial perturbations at the magnitude of $O(10^{-4})$ are randomly generated. The red circle indicates the non-Quincke region satisfying \refeqq{quincke3}. The markers p1 to p4 indicate four positions when $z$ first hits the value of $0.0,1.0,2.0,3.0$. $G=0.4000$. The particle does not rotate in the equilibrium state. }
\label{dynfig}
\end{figure}

\begin{figure}[h!]
{\centerline{\includegraphics[width=3.8in]{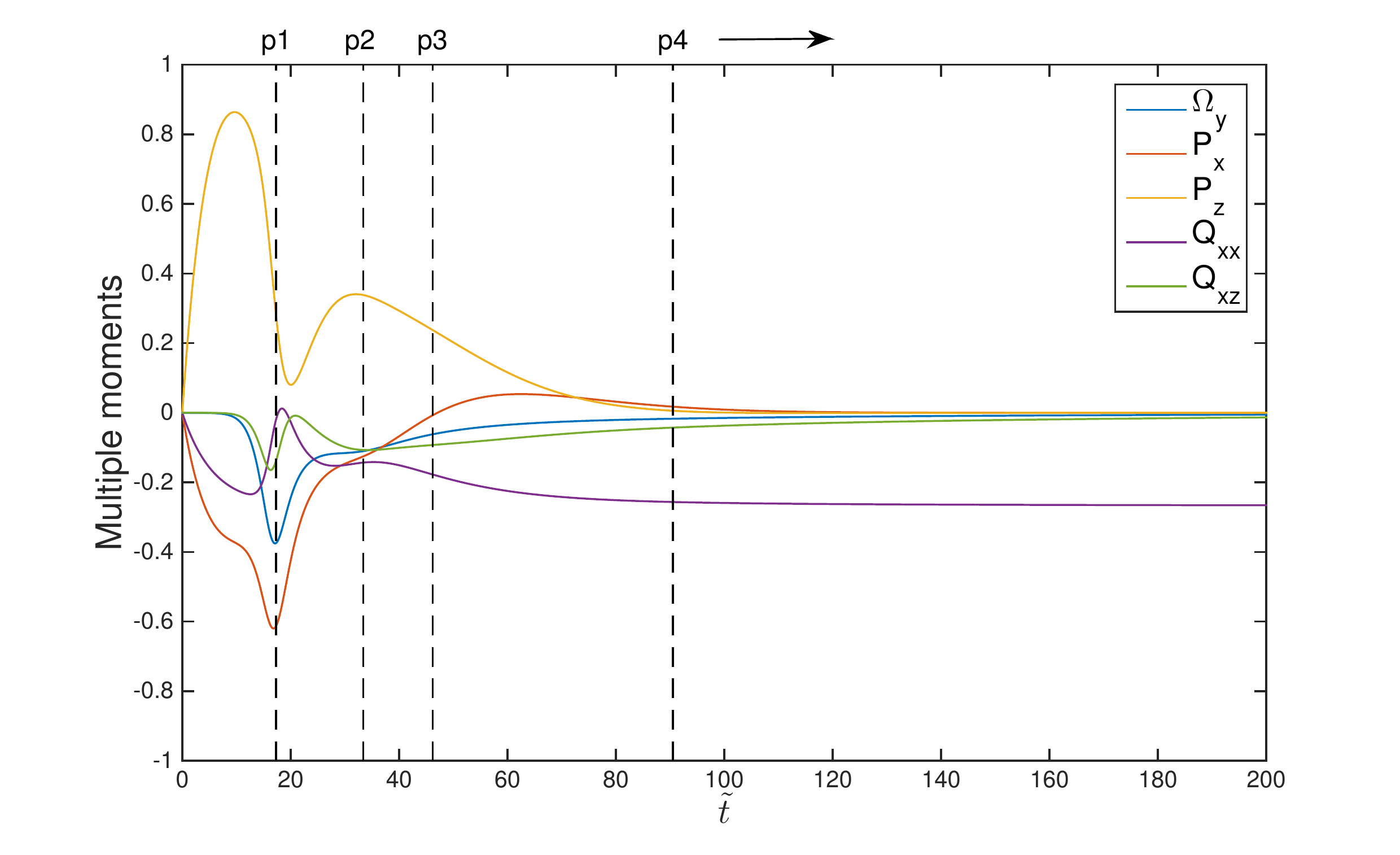}}}
\caption{\footnotesize  The evolution of corresponding multiple moments and rotation rate to the single particle dynamic in Fig \ref{dynfig}.}
\label{sigdyn2}
\end{figure}

Figure \ref{dynfig} shows a typical  particle trajectory when $G^*=0.4000$. The particle undergoes negative dielectrophoresis (DEP) and moves  towards the minimum field location. The Quincke rotation and continuous changing DEP force make the trajectory non-straight 
The circle in Figure \ref{dynfig} indicates the region given by \refeqq{quincke3}. Within this region, the particle does not undergo Quincke rotation. The evolution of particle rotation rate and the components of the dipole and quadrupole moments is shown in Figure \ref{sigdyn2}. 
Upon particle release,  the magnitude of its rotation rate increases. Position p1 is a turning point after which the rotation magnitude starts to decrease. Positions p2 and p3 are two intermediate points before particle enters the 'non-Quincke' zone. Position p4 is when the particle arrives at the center axis, where rotation and all multiple moments except the diagonal elements of $\bQ$ decay to zero.

Increasing of the field gradient strength shrinks the `non-Quincke' zone and when $G\geq G^*$ the electro-rotation occurs everywhere in the space. A single particle will eventually stay steady at the equilibrium position with a non-zero rotation rate.

\begin{figure*}
\includegraphics[width=6.5in,height=3.8in]{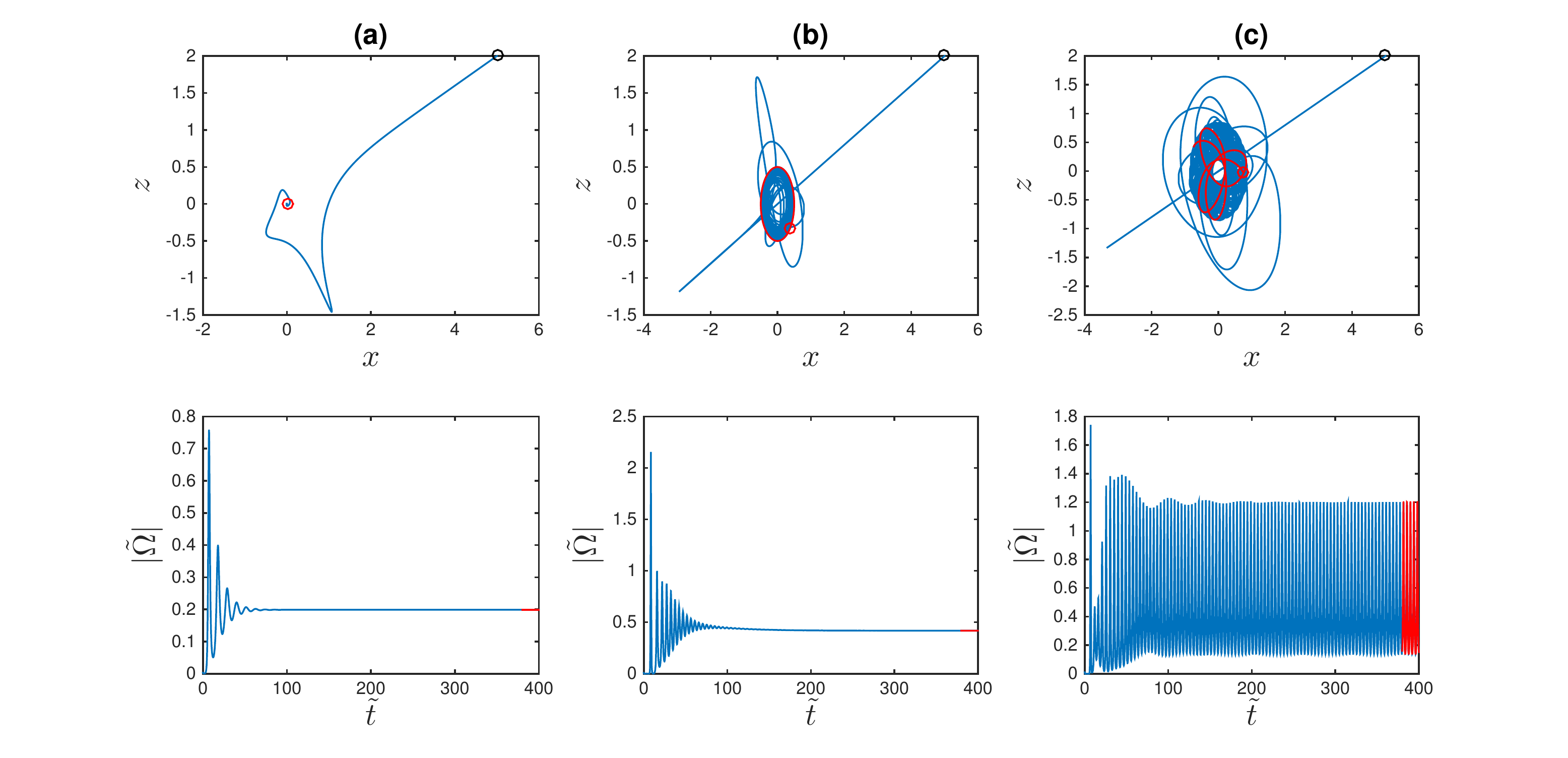}
\caption{\footnotesize  Particle trajectories in different field gradient strength. Initial position $x=5.0$, $z=2.0$ and a random initial polarization perturbation at $O(10^{-4})$. $D=5.1520$, $D'=5.6054$. (a). $G=1.0,$ (b).  $G=2.3,$ (c). $G=3.0.$ The final time interval $\hat t=380 ~400$ is indicated by red color. See supplementary material for movies \cite{sm}.}
\label{bifurfig}
\end{figure*}

In even stronger fields, particle dynamics becomes more complex.
 Figure \ref{bifurfig} shows particle steady state changing from a stable point (Fig \ref{bifurfig}(a)), to circular orbit (Fig \ref{bifurfig}(b)) and finally bounded elliptic orbit (Fig \ref{bifurfig}(c)) as $G$ increases.

\begin{figure}[h]
\centerline{\includegraphics[scale=0.4]{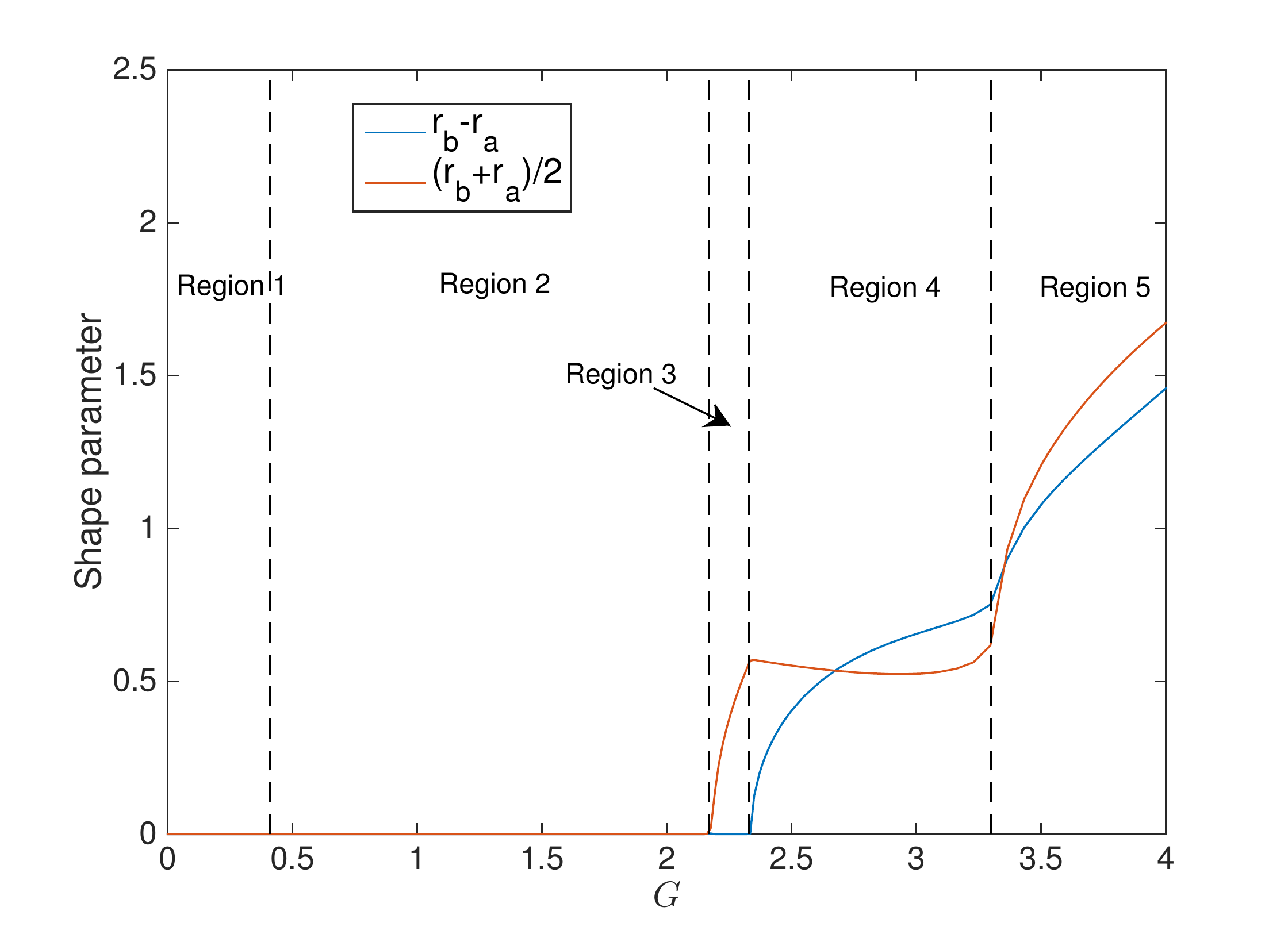}}
\caption{\footnotesize  Stationary trajectory shape vs $G$. $r_a$ and $r_b$ are the short and long radius of the steady orbit. When no steady orbit is observed, $r_a$ and $r_b$ are the minimum and maximum distance to the origin in a chosen time frame. $D'=5.6054$, $\epsilon'_{mw}=-0.0670$, $\sigma'_{mw}=-0.3333$. }
\label{bifr}
\end{figure}

The longtime stationary trajectory can be characterized by  the short  $r_a$ and the long $r_b$ axes of the elliptical orbit. Figure \ref{bifr} shows the dependence of  $r_a$ and $r_b$ on field strength. Region 1 (no rotation) and 2 indicate steady position in the physical plane ($r_a=r_b=0$). In region 2, the particle undergoes electro-rotation, however there is no  off-center particle motion in this steady state solution(see Fig \ref{bifurfig}(a)). The narrow region 3 corresponds to a different scenario(see Fig \ref{bifurfig}(b)). In this region, the origin is no longer a steady position. The particle trajectory converges to a circle centered at the origin. The radius of this circle increases with $G$. Region 4 shows a transition from the circular trajectory to  rotating elliptical trajectories as in Fig \ref{bifurfig}(c). It is interesting to notice that in this region, the average radius stays almost unchanged, but the deformation increases with $G$. Region 5 indicates a simultaneous increase of the deformation and the average orbit radius, while the orbit retains elliptical shape.

\section{Multi-particle System}
Here we extend \refeqq{p1}, \refeqq{qc1} and \refeqq{mt1} to many particles systems. For convenience, we revert to dimensional variables.

\subsection{General formalism}
Here we generalize the model to describe the dynamics of  N particles. With respect to the evolution of multipole moments in \refeqq{p1} and \refeqq{qc1}, now we need to include the disturbance from other particles. Assuming widely separated particles, introduce the two vectors $\bE^p_{ij}=\nabla\frac{\br\cdot \bP_j}{r^3}\big |_{\br=\bR_{ij}}$ and $\bE^q_{ij}=\nabla\frac{\br\br: \bQ_j}{2r^5}\big |_{\br=\bR_{ij}}$, as the two leading orders of the field contributions valued at particle $i$ from the expansion of distrubance potential induced by particle $j$. Also introduce $\bG^p_{ij}=\nabla\nabla\frac{\br\cdot \bP_j}{r^3}\big |_{\br=\bR_{ij}}$. Here $\mathbf{R}_{ij}=\mathbf{r}_{j}-\mathbf{r}_{i}$.

The evolution equations of the multipole moments of the $i^{th}$ particle are 
\begin{equation}
\begin{split}
\label{multiP}
\frac{d \bP_i}{dt}=&\bOmg_i\times [\bP_i+a^3\epsilon_{cm}(\nabla\phi_a(\br_i)+\sum_{j\neq i} (\bE^p_{ij}+\bE^q_{ij}))]\\
&-\frac{1}{\tau_{mw}}[\bP_i+a^3\sigma_{cm}(\nabla\phi_a(\br_i)+\sum_{j\neq i} (\bE^p_{ij}+\bE^q_{ij}))],
\end{split}
\end{equation}
and

\begin{equation}
\begin{split}
\label{multiQ}
\frac{d \bQ_i}{dt}=&\{\bOmg_i \times  [\bQ_i+2a^5\epsilon'_{cm}(\nabla\nabla\phi_a(\br_i)+\sum_{j\neq i} \bG^p_{ij}]\}^{sym}\\
&-\frac{1}{\tau'_{mw}}[\bQ_i+2a^5\sigma'_{cm}(\nabla\nabla\phi_a(\br_i)+\sum_{j\neq i} \bG^p_{ij}],
\end{split}
\end{equation}
where $sym$ denotes $A_{ij}^{sym}=A_{ij}+A_{ij}\tran$. 
The truncation error is $O((a/R_{ij})^{-5})$ in this approximation \cite{Hu2017}. Here it is assumed that the minimum $R_{ij}=|\bR_{ij}|$ over all $i\neq j$ is used to estimate the error.

The electric force on a particle is (see Appendix B for details)
\begin{equation}
	\bF_i^{el}=\bF_i^{d1}+\bF_i^{d2}+\bF_i^{d3},
\end{equation}
where 
\begin{equation}
\begin{split}
	\bF_i^{d1}=&4\pi\epsilon_f \bP_i\cdot \nabla\nabla \phi_a(\mathbf{r}_i),\\
	\bF_i^{d2}=&-\sum_{j\neq i}\frac{12\pi\epsilon_f}{R^4_{ij}}[(\bP_i\cdot \widehat{\bR}_{ij})\bP_j+(\bP_j\cdot\widehat{\bR}_{ij})\bP_i\\
	&+(\bP_i\cdot \bP_j)\widehat{\bR}_{ij}-5(\bP_j\cdot\widehat{\bR}_{ij})(\bP_i\cdot\widehat{\bR}_{ij})\widehat{\bR}_{ij}],\\
	\bF_i^{d3}=&\frac{2\pi\epsilon_f}{3} \bQ_i:\nabla\nabla\nabla\phi_a(\mathbf{r}_i).
\end{split}
\end{equation}
The electric torque is
\begin{equation}
	\bT_i^{el}=\bT_i^{d1}+\bT_i^{d2}+\bT_i^{d3}+\bT_i^{d4},
\end{equation}
where
\begin{equation}
\begin{split}
	\bT_i^{d1}=&4\pi\epsilon_f (\bP_i\times \nabla\phi_{a}(\mathbf{r}_i)+(\bQ_i\cdot \nabla)\times\nabla\phi_{a}(\br_i)),\\
	\bT_i^{d2}=&-4\pi\epsilon_f \bP_i\times \sum_{j\neq i}\bE^p_{ij},\\
	\bT_i^{d3}=&-4\pi\epsilon_f \bP_i\times \sum_{j\neq i}\bE^q_{ij},\\
	\bT_i^{d4}=&-4\pi\epsilon_f(\bQ_i\cdot \nabla)\times(\sum_{j\neq i}\bE^p_{ij}).
\end{split}
\end{equation}

The quadrupole-quadrupole interaction and higher order moments are neglected since they come at the order of $O((a/R_{ij})^{-5})$, as shown in Appendix B. 

Particle motion generates fluid flow, hence hydrodynamic interactions should also be taken into account \cite{Kim-Karrila:1991}. 
Including the particles-induced flow, the equations of motion of a particle are
\begin{equation}
\begin{split}
\label{ueqn}
	\bu_i&=\frac{\bF^{el}_i}{6\pi a\eta_f}+\sum_{j\neq i}\frac{\bF^{rep}_{ij}}{6\pi a\eta_f}+\sum_{j\neq i}\frac{(5\bF^{el}_i\cdot\widehat{\bR}_{ij})\widehat{\bR}_{ij}}{8\pi a\eta_f R^4_{ij}}\\
	&+\frac{1}{\eta_f}\sum_{j\neq i}[-\frac{\bT_j^{el}\times \widehat{\bR}_{ij}}{8\pi R^2_{ij}}\\
	&+\frac{1}{8\pi}(\frac{1}{R_{ij}}+\frac{2a^2}{3R^3_{ij}})\bF^{el}_j\\
	&+\frac{1}{8\pi}(\frac{1}{R_{ij}}-\frac{2a^2}{R^3_{ij}})(\bF^{el}_j\cdot\widehat{\bR}_{ij})\widehat{\bR}_{ij}],
	\end{split}
\end{equation} 
	\begin{equation}
\begin{split}
\label{oeqn}
	\bOmg_i&=\frac{\bT_i^{el}}{8\pi a^3\eta_f}\\
	&+\frac{1}{\eta_f}\sum_{j\neq i}[-\frac{\bT_j^{el}}{16\pi R^3_{ij}}-\frac{3}{16\pi R^3_{ij}}(\bT_j^{el}\cdot\widehat{\bR}_{ij})\widehat{\bR}_{ij}\\
	&-\frac{(\bF^{el}_j) \times \widehat{\bR}_{ij}}{8\pi R^2_{ij}}].
\end{split}
\end{equation}

In \refeqq{ueqn}, we introduce an artificial isotropic repulsion force \cite{Yeo:2010b} to prevent particle contact,
\begin{equation}
\label{rep}
\bF^{rep}_{ij}=F^r_0(\frac{r^2_c-|\bR^2_{ij}|}{r^2_c-4a^2})^2\widehat{\bR}_{ij}, \quad R_{ij}<r_c,
\end{equation}
where $r_c=2.01a$ is a control distance used to simulate surface roughness and $F^r_0$ is a characteristic repulsion force unit.

The detailed derivation of \refeqq{ueqn} and \refeqq{oeqn} is provided in the Appendix B. For an applied linear electric field, the evolution equations for the multipole moments are exact.
For a general non-uniform electric field, especially a field which rapidly varies on the particle scale, the truncation of multipole moments as well as the Taylor expansion of the applied field introduce error. However, we are still able to set up a similar model for slowly varying fields by doing a systematic asymptotic analysis and assuming a proper balancing order for the scales of the applied field and the disturbance field. The discussion is provided in the Appendix C.

\subsection{Two-particle dynamics}

\begin{figure*}
\centering
\includegraphics[width=7.5in,height=4.0in]{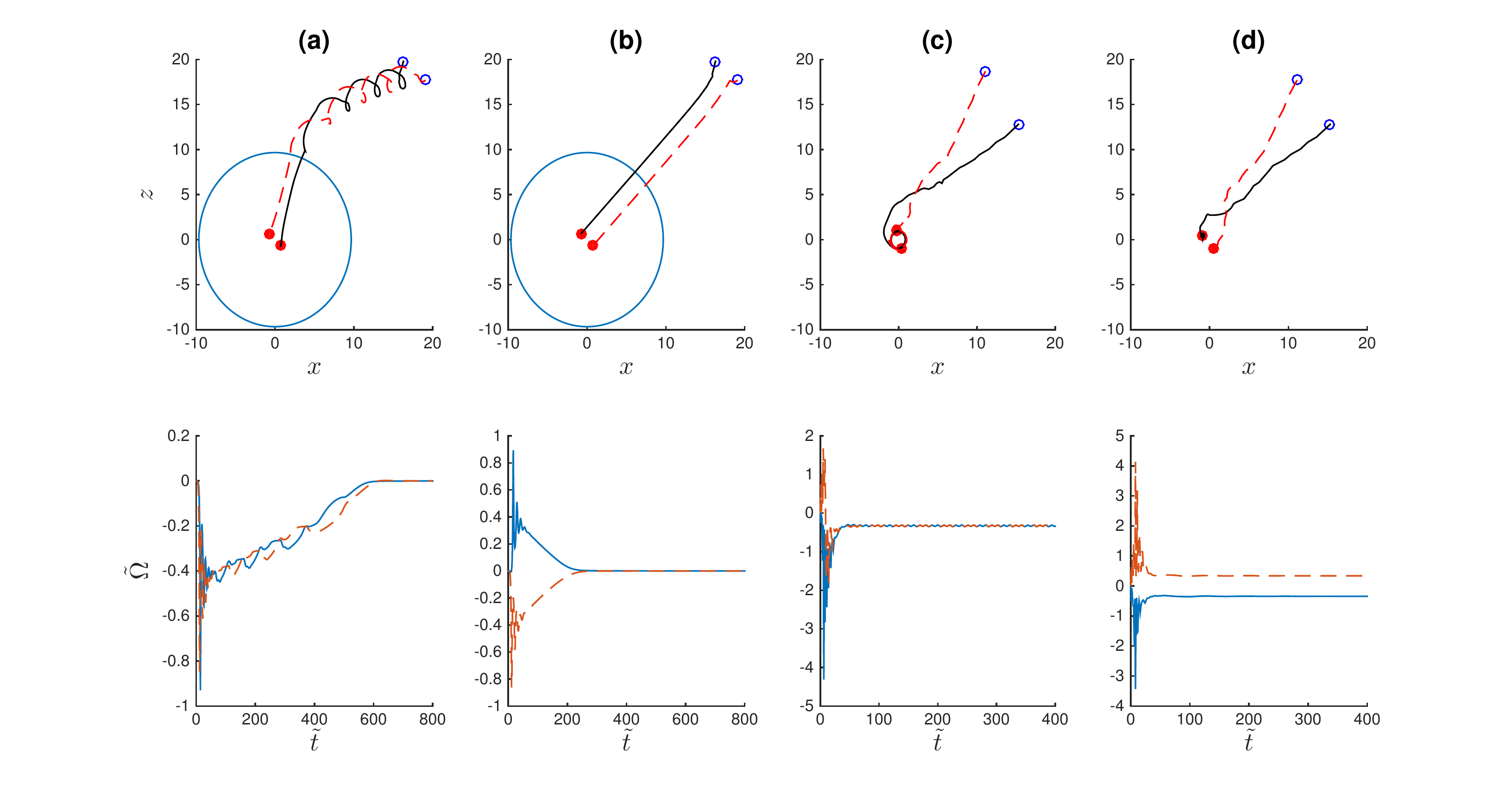}
\caption{\footnotesize  Particle dynamic patterns and their rotation rates as a function of time. The dash(solid) line in the upper figure corresponds to the dash(solid) line in the lower. Initial $\bP$ is given randomly at $O(10^{-4})$  The non-electrorotation region is indicated by the circle satisfying \refeqq{quincke3}.  $D=5.1520,$ $D'=5.6054$. (a) co-rotating pair, $G=0.1$, (b) counter-rotating pair, $G=0.1$, (c) co-rotating pair, $G=1.0$, (d) counter-rotating pair, $G=1.0$. See supplementary material for movies \cite{sm}.}
\label{pairdyn}
\end{figure*}

Figure  \ref{pairdyn} shows the interaction of two identical particles at different applied field strenghts $E_0$.  The spheres move towards the origin (location of minimum field), due to the dielectophoretic (DEP) force, while also executing rotations, due to the Quincke effect.
In a uniform field, the spheres would orbit around each other \cite{Das-Saintillan:2013, Lushi-Vlahovska:2014}. In the non-uniform field, this orbiting motion is superimposed on the DEP translation.  The circle which satisfies \refeqq{quincke3} is drawn in Figure \ref{pairdyn}.(a)-(b). This circle represents the boundary of the existence of a steady nonzero $\bOmg$, hence within the circle only transient rotation can exist.

In our computations, the two particles are positioned in the electrorotation region and random initial polarizations are applied. Computations for different initial polarizations and different $G$ are presented in Figure \ref{pairdyn}.  In Figure \ref{pairdyn}.(a), the random initial polarizations have the particles initially rotating in the same direction, and hydrodynamic interactions then drive the particle pair to orbit about each other. Meanwhile they translate towards the `non-Quincke' region due to the DEP force. The rotation decays to zero once they enter it. In Figure \ref{pairdyn}.(b),  the spheres are initially counterrotating  and form a translating pair moving quite linearly to the `non-Quincke' region. 
In stronger fields, the non-rotation region shrinks. As shown in Figure \ref{pairdyn}.(c)-(d), the DEP force from the external field dominates the pair interactions. However when particles come close to each other, we observe a pairing phenomenon due to their rotation. In the first example, Figure \ref{pairdyn}.(c), the two particles form a co-rotating cluster. In Figure \ref{pairdyn}.(d), we find that the two particles form a stationary counter-rotating pair due to the balance of DEP force and hydrodynamic interaction.

We note that in the previous pair-particle cases, the particles are initially in the $ x-z$ plane of the applied electric field and no initial disturbance is given in the $y$-direction 
Thus no motion in the $y$-direction is present. However, if the particle's initial alignment is not in the $x-z$ plane of the electric field or there is any orthogonal perturbation (in the $y$-direction ), the in-plane motion is not stable. In this case particles eventually form a chain orthogonal to the field plane, i.e.  along y-axis in our field set-up. The particle's axis of rotation is then orthogonal to the $x-z$ plane. This observation will be applied to the multi-particle case below.

\subsection{Multi-particle dynamics in nonuniform fields}

The nonuniform electric field can be utilized to assemble structures of spheres.

\begin{figure}[h!]
\centering
  \subfloat[$\widetilde t=0$]{%
    \includegraphics[width=.22\textwidth,height=.2\textwidth]{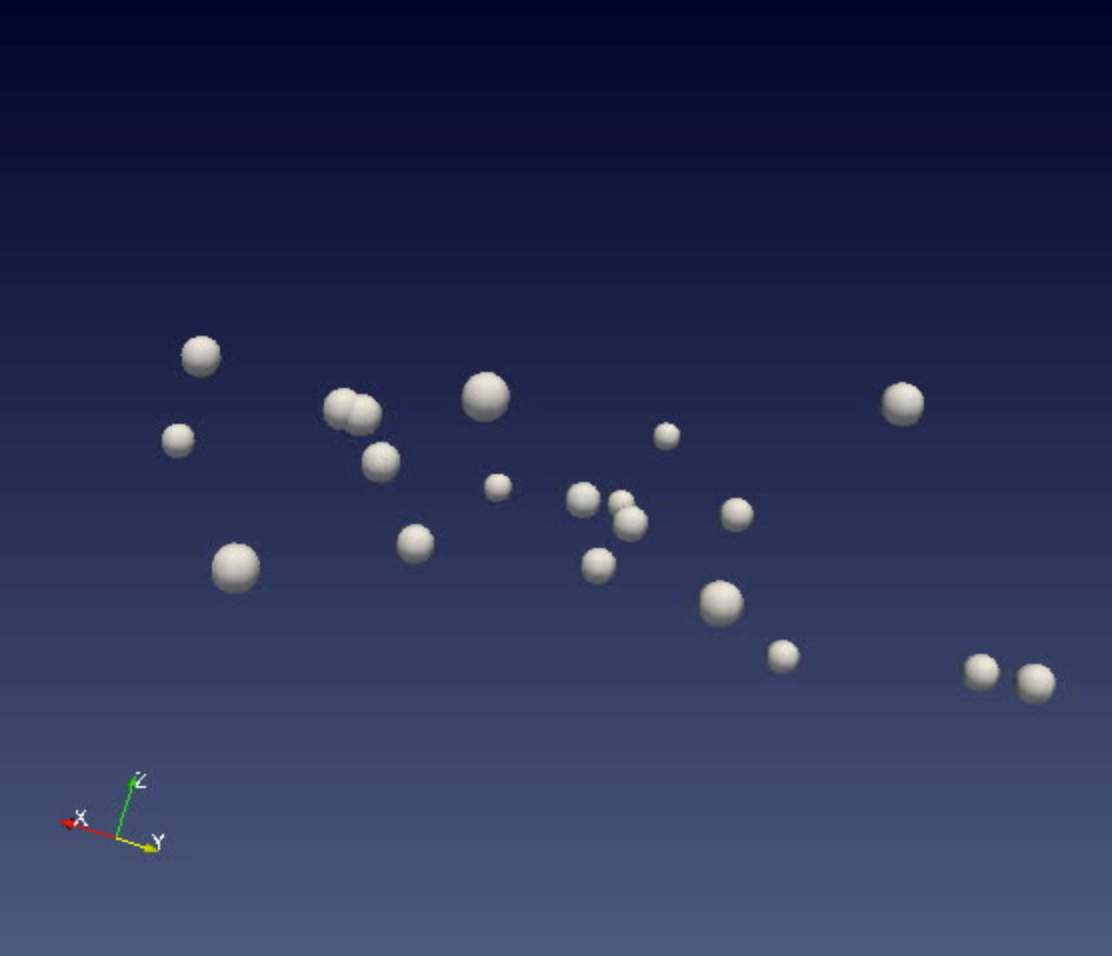}}\hfill
  \subfloat[$\widetilde t=20$]{%
    \includegraphics[width=.22\textwidth,height=.2\textwidth]{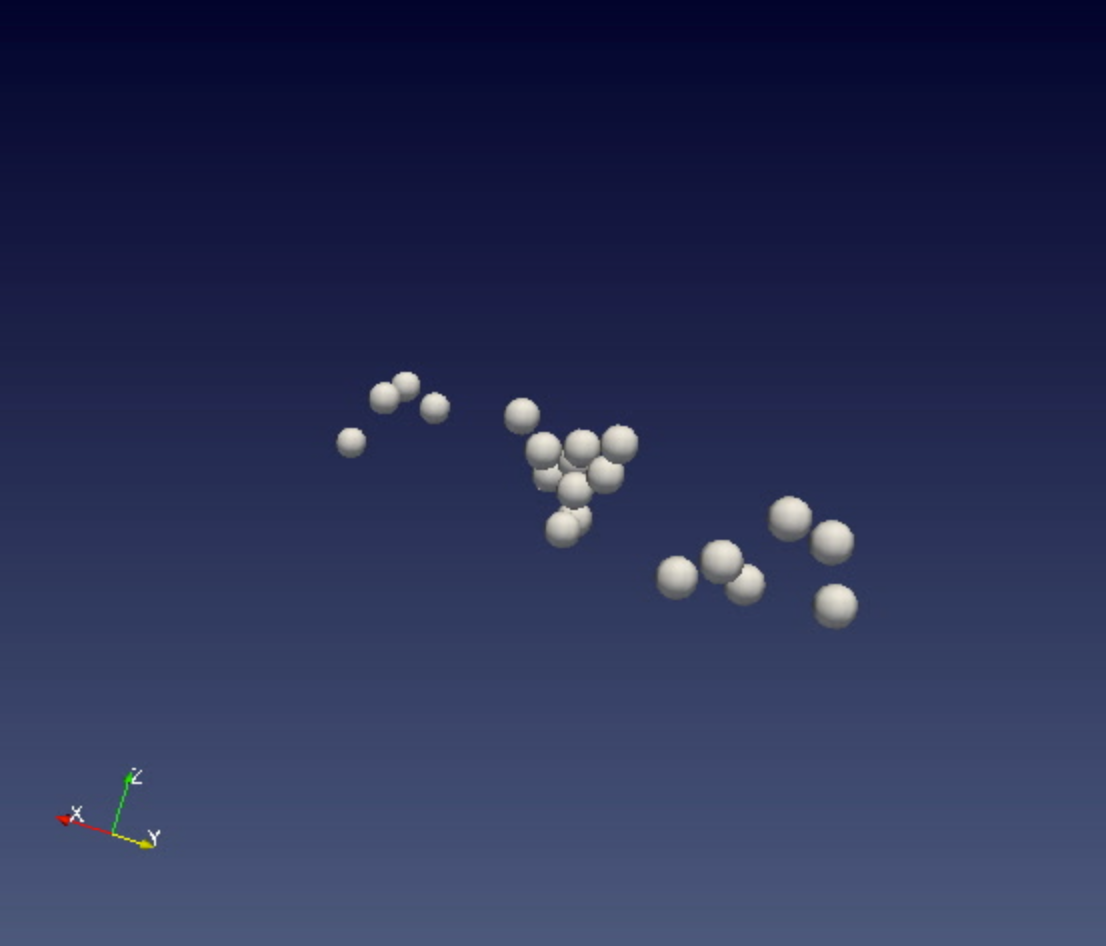}}\hfill
  \subfloat[$\widetilde t=200$]{%
    \includegraphics[width=.22\textwidth,height=.2\textwidth]{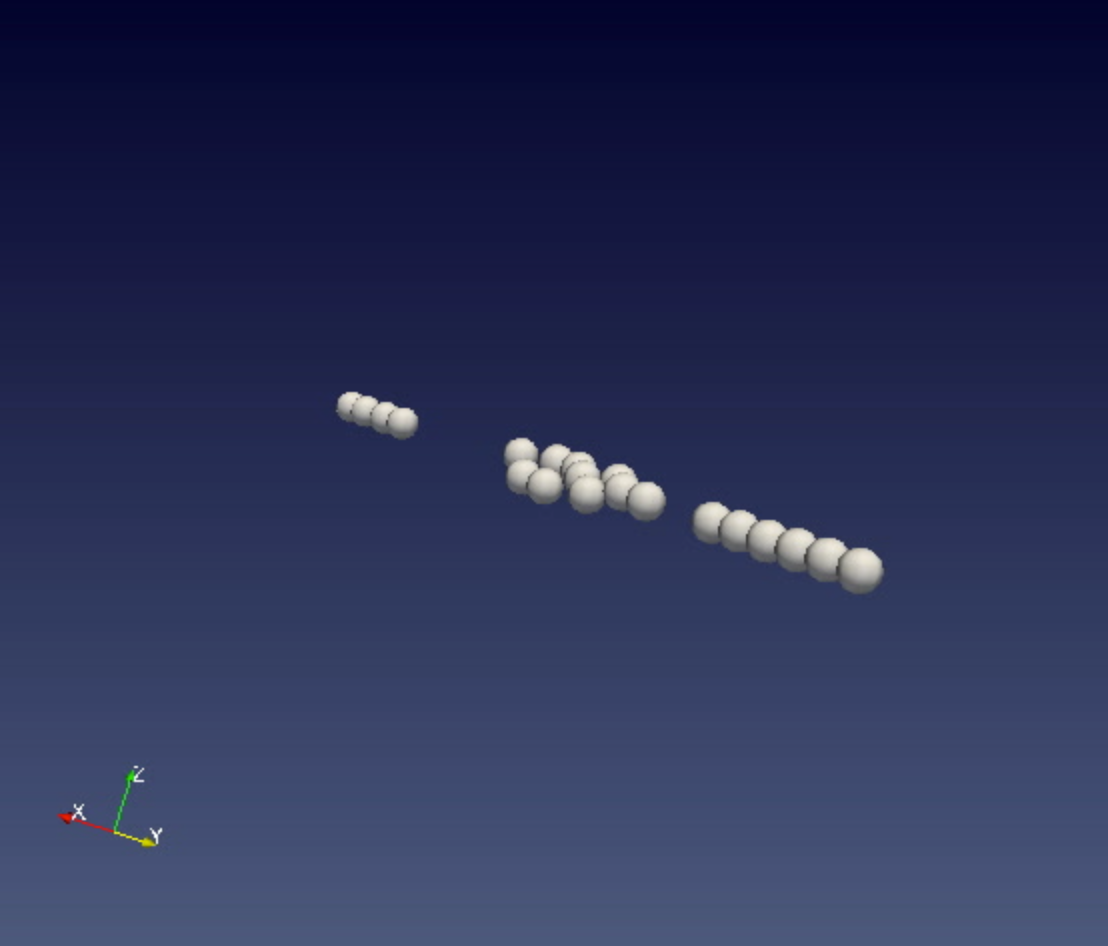}}\hfill
  \subfloat[$\widetilde t=2000$]{%
    \includegraphics[width=.22\textwidth,height=.2\textwidth]{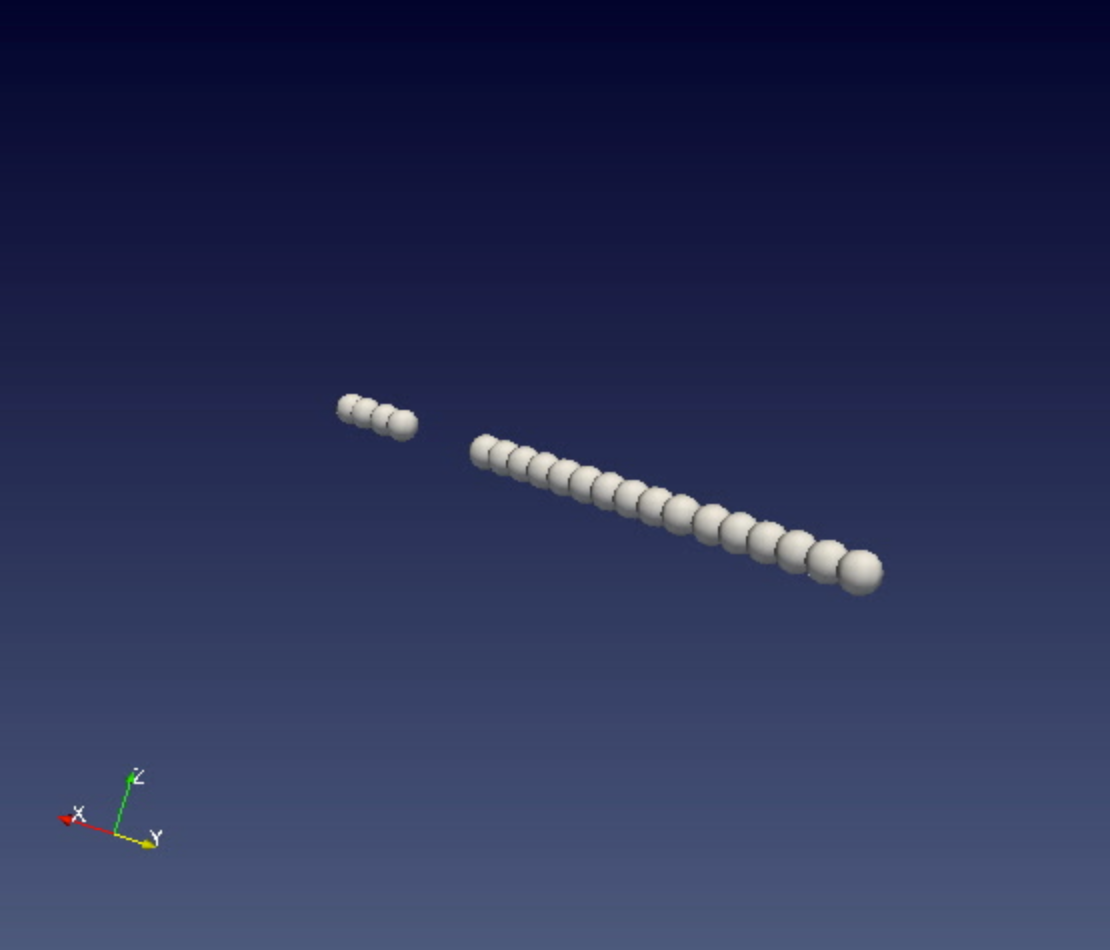}}
  \caption{\footnotesize  Dynamics of 20 particles in a linear field with random initial positions. Chaining at Y-axis is observed.  $G=1.0.$ $D=5.1520.$ $\widetilde t=0, 20, 200, 2000.$ See supplementary material for movies \cite{sm}.}\label{multifig}
\end{figure}

 Figure \ref{multifig} shows that spheres in a linear field \refeqq{eap} form a chain  extending along the $y$ direction, which is the direction of rotation.
However in stronger fields, no stable assembly exists and the multi-particle dynamics is complex (similar to the single particle scenario).

\begin{figure}[h!]
\centering
  \subfloat[$\widetilde t=0$]{%
    \includegraphics[width=.22\textwidth,height=.2\textwidth]{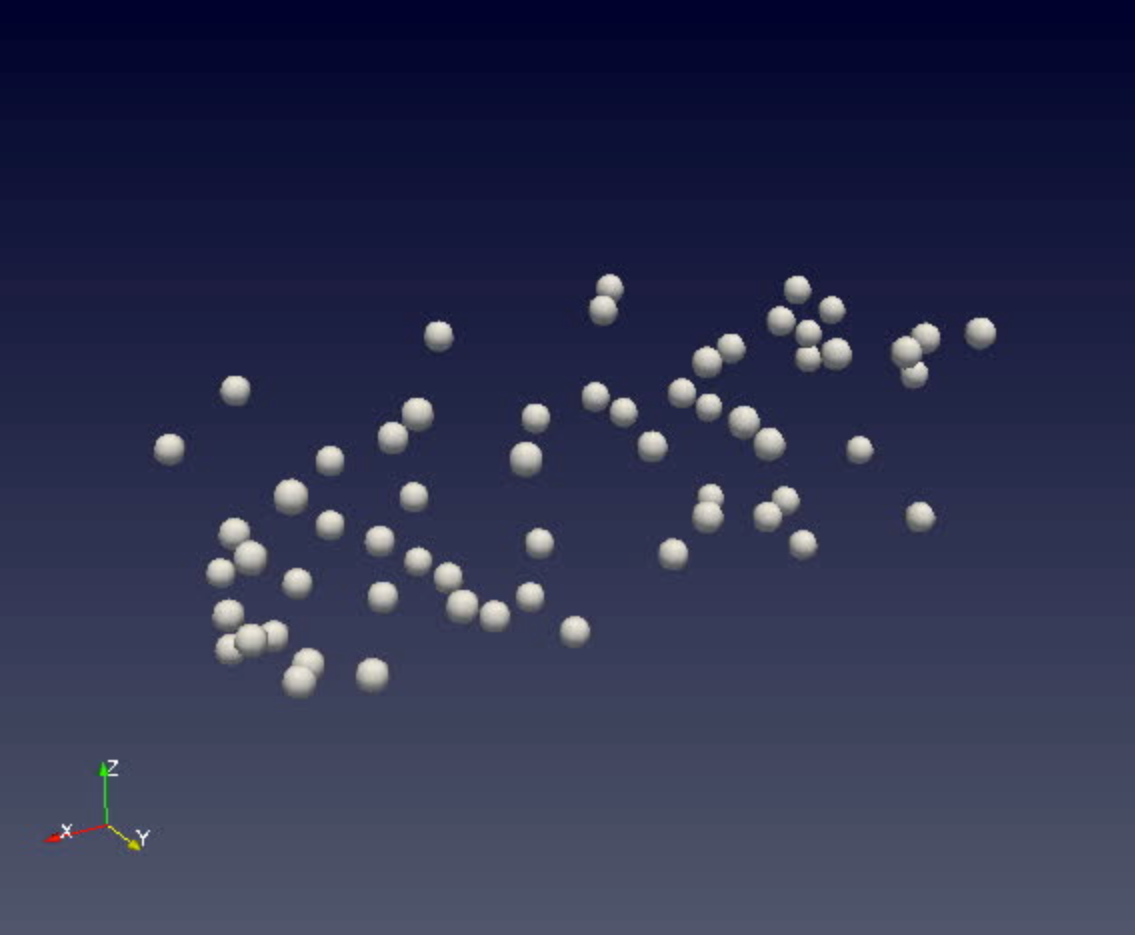}}\hfill
  \subfloat[$\widetilde t=20$]{%
    \includegraphics[width=.22\textwidth,height=.2\textwidth]{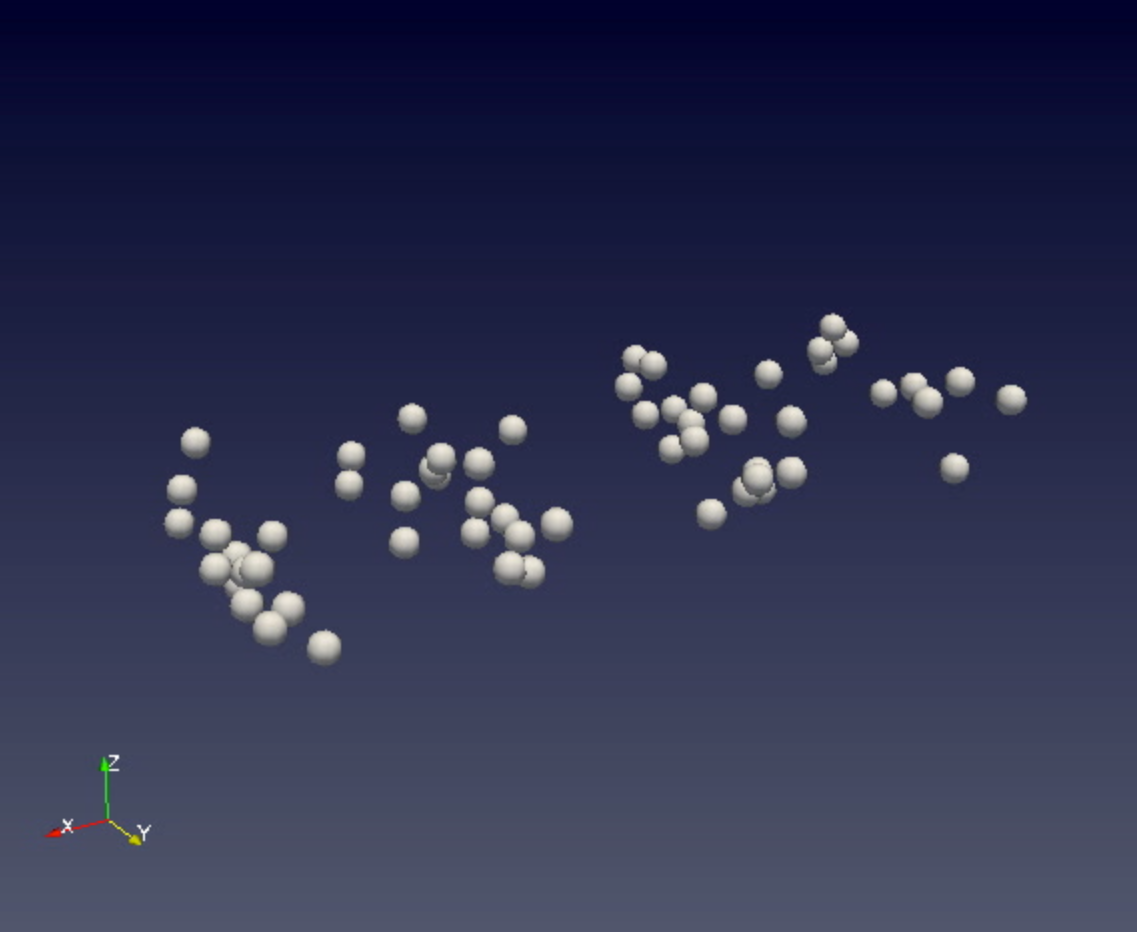}}\hfill
  \subfloat[$\widetilde t=200$]{%
    \includegraphics[width=.22\textwidth,height=.2\textwidth]{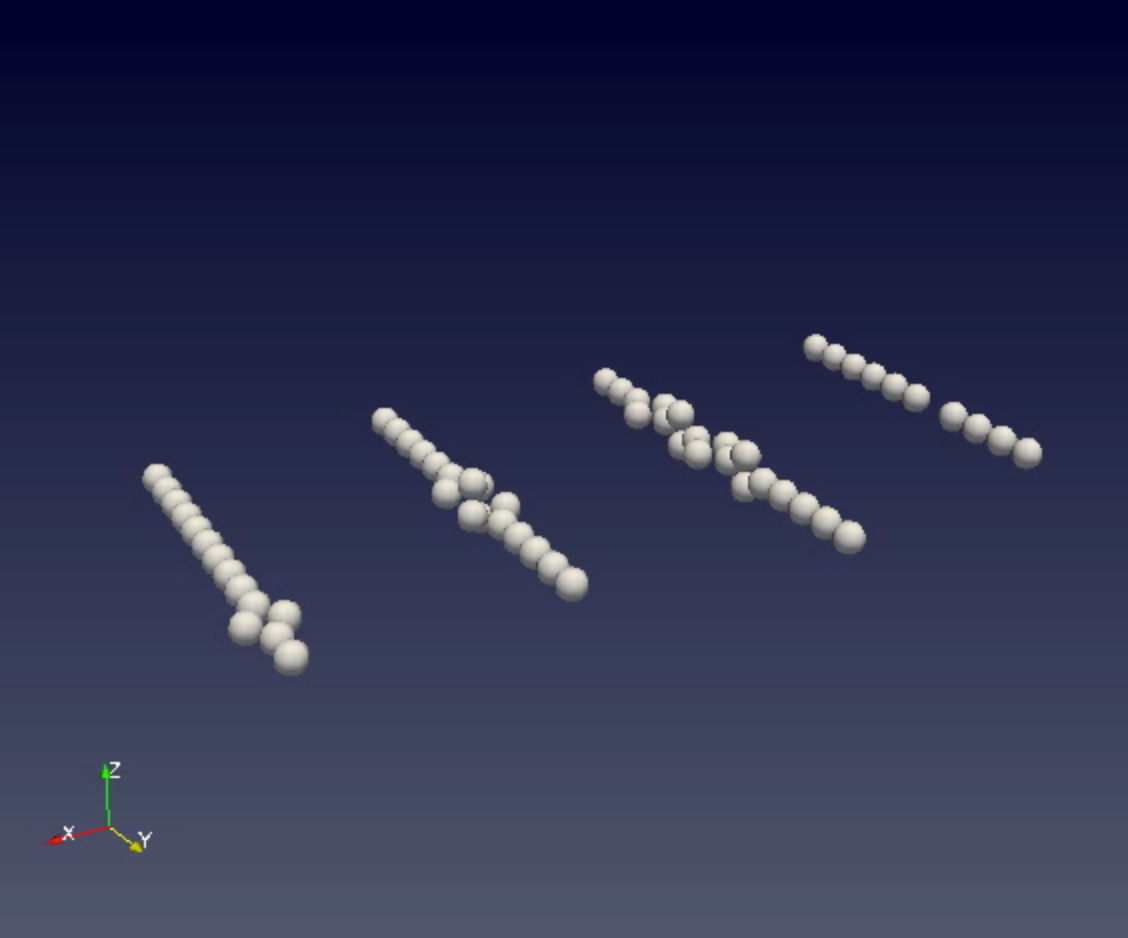}}\hfill
  \subfloat[$\widetilde t=2000$]{%
    \includegraphics[width=.22\textwidth,height=.2\textwidth]{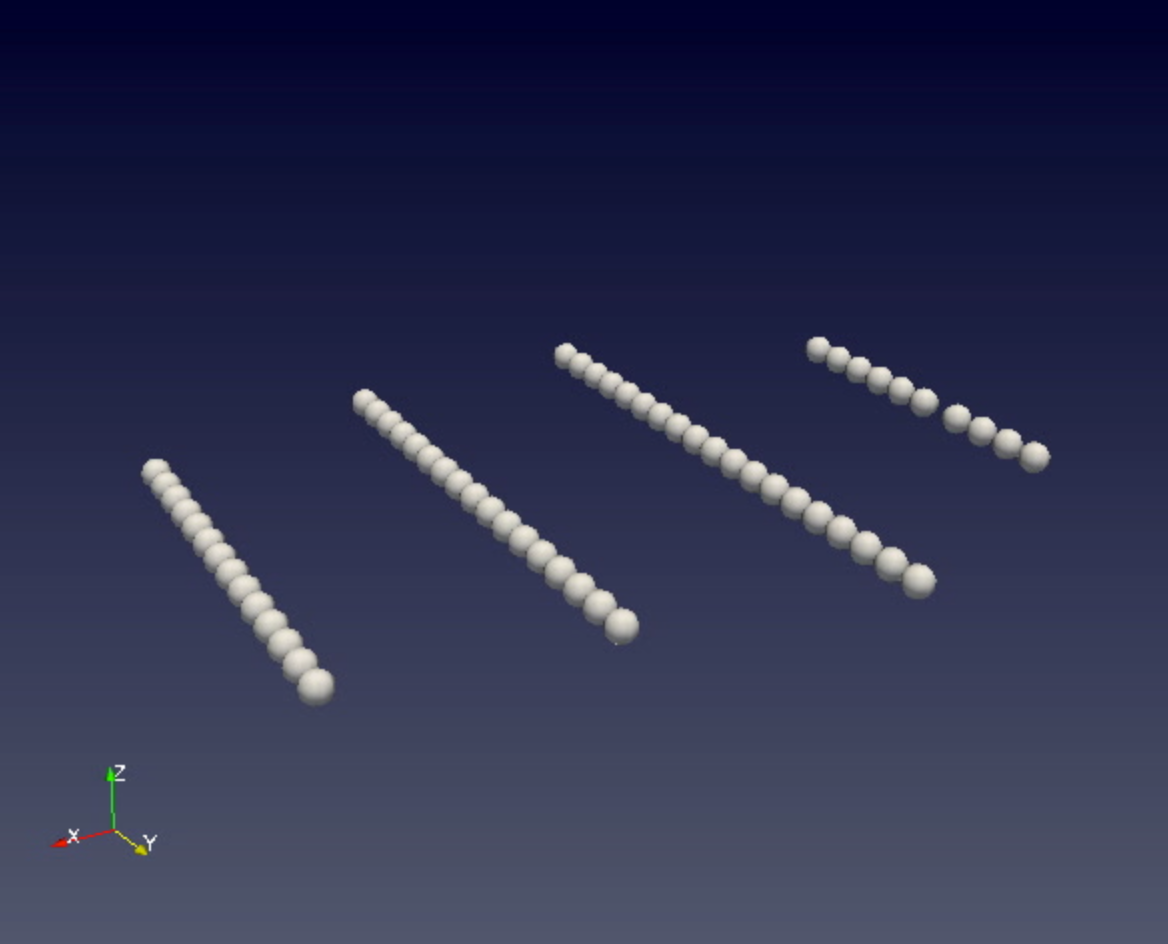}}
  \caption{\footnotesize  Dynamics of 60 particles in periodical field with random initial positions with $\tilde x \in [0,60]$. Clustering at $\tilde z=0$ plane is observed. Four chains are formed at $\tilde x=8,24,40,56.$ $E_0/E_c=1.0.$ $\delta'=\frac{\pi}{16}$. $D=5.1520.$ $\widetilde t=0, 20, 200, 2000.$ See supplementary material for movies \cite{sm}.}\label{sinfig}
\end{figure}

Our multi-particle  model can be applied to study particle dynamics in a  more general non-uniform fields. 
Here we illustrate particles assembly in a slowly varying periodical electric potential which generates a spatially-periodic electric field
\begin{equation}
\tbE_{a}=\frac{E_0}{E_c}[\delta'\sin(\delta' \tilde x)\sinh(\delta' \tilde z){\bf{i}}-\delta'\cos(\delta' \tilde x)\cosh(\delta' \tilde z){\bf{k}}].	
\end{equation}
The field is periodic in $x$ direction and can be generated by two separated plain electrodes with opposite sinusoidal applied potential. 
Figure \ref{sinfig} illustrates the particle configurations for the case  $E_0=E_c$ and $\delta'=\frac{\pi}{16}$  and 60 particles. 
The particles are observed to cluster in the $x-y$ plane and form chains periodically localized at all zero points of field strength. This is also majorly due to a negative DEP effect. In our simulations the chains' positions are eventually stable while each particle undergoes steady rotation.

\section{Conclusions}


We developed a model to investigate the dynamics of  spheres in a nonuniform electric field when Quincke rotation has significant effect on the dynamics. Our theory is built on the Taylor-Melcher leaky dielectric model, which assumes ohmic conduction in the bulk and creeping flow. Considering an applied field with spatial variations much larger than the inter-particle spacing and the radius of the spheres, particle polarization is approximated by the dipole and quadrupole moments. This reduces the problem to a system of ordinary differential equations for the particle position, rotation rate,  dipole and quadrupole moments. In this paper we focus on an applied linear electric field in which case the approximation is exact.

In the study of the steady state of single sphere, we identify a necessary condition for when the nonuniform field induces Quincke rotation. We find the threshold for electorotation in a linear field. Increasing the electric field strength, makes the particle dynamics more complex: while Quincke electrorotation is characterized by  steady spinning around the particle  center, in stronger fields  time-dependent orbiting motion around the minimum field location is observed.


We generalize the model to consider multi-particle simulations in arbitrary (but  spatially slowly varying) applied fields. Hydrodynamic interactions are included via the grand-mobility matrix. In the leading order analysis, we retain terms in the far field expansions up to forth order in the inter particle spacing. 
The electrostatic interactions between particles include dipole-dipole and dipole-quadrupole interactions which are naturally introduced from the dielecropheritc force calculation.  A numerical study of two-particle and multi-particle systems were considered for the special case of a linear applied field. Our calculation show intricate trajectories  in the case of pairs, and chain-like assemblies in the case of many particles. 


 Our model provides a basis to study the collective  dynamics of many particles in a general  electric field. 
 The model can also be extended to include ambient flow,  and thus can be  applied to study problems in electrorheology.

\section{Acknowledgment}

This work was supported in part by NSF grant DMS-1312935 and 1716114, and CBET-1437545 and 1704996.

\appendix

\section{Deriving evolutions of multipole moments}
The Taylor expansion of the external field at the particle center is given by,
\begin{equation}
\label{apt2}
\begin{split}
	\phi_a(\mathbf{r})=&\phi_a(\bzero)+\br\cdot \nabla\phi_a(\bzero)+\frac{1}{2}\br\br: \nabla\nabla\phi_a(\bzero)\\
	&+\frac{1}{6}\br\br\br[\cdot]^3\nabla\nabla\nabla\phi_a(\bzero)+\cdots.
\end{split}
\end{equation} 
Then the gradient of potential is,
\begin{equation}
\label{apt2g}
	\nabla \phi_a(\mathbf{r})=\nabla\phi_a(\bzero)+\nabla\nabla\phi_a(\bzero)\cdot \br+\frac{1}{2}\nabla\nabla\nabla\phi_a(\bzero):\br\br+\cdots.
\end{equation} 
	
As we discussed in Sec. II.A, the induced potential due to the particle could be expanded in a series of spherical harmonics as,	
\begin{equation}
\label{apt3}
\begin{split}
	\phi_d(\mathbf{r})=\frac{\mathbf{r\cdot P}}{|\br|^3}+\frac{1}{2}\frac{\mathbf{rr: Q}}{|\br|^5}+\frac{1}{2}\frac{\br\br\br[\cdot]^3\bO}{|\br|^7}+\cdots, \qquad |r|>a,\\
	\bar \phi_d(\mathbf{r})=\frac{\mathbf{r\cdot P}}{a^3}+\frac{1}{2}\frac{\mathbf{rr: Q}}{a^5}+\frac{1}{2}\frac{\br\br\br\tvdots\bO}{a^7}+\cdots, \qquad |r|\leq a,
\end{split}
\end{equation} 
where $\bP$, $\bQ$ and $\bO$ are the dipole, quadrupole and octopole moments.

We obtain the gradients as,
\begin{equation}
\label{apt3g}
\begin{split}
	\nabla\phi_d(\mathbf{r})&=\frac{\bP}{|\br|^3}-\frac{3\mathbf{r\cdot P}}{|\br|^5}\mathbf{r}+\frac{1}{2}(\frac{2\mathbf{Q\cdot r}}{|\br|^5}-\frac{5\mathbf{rr: Q}}{|\br|^7}\mathbf{r})\\
	&+\frac{1}{2}(\frac{3\bO:\br\br}{|\br|^7}-\frac{7\bO\tvdots\br\br\br}{|\br|^9}\mathbf{r})+\cdots, \qquad |\br|>a,\\
	\nabla \bar \phi_d(\mathbf{r})&=\frac{\mathbf{ P}}{a^3}+\frac{\mathbf{Q\cdot r}}{a^5}+\frac{3}{2}\frac{\bO:\br\br}{a^7}+\cdots, \qquad |\br|\leq a.
\end{split}
\end{equation}

We substitute the above gradients into the charge conservation equation, \refeqq{qc}.
From the definition of the induced charge $q$ we obtain,
\begin{equation}
\label{aq}
\begin{split}
	q&=-\mathbf{n}\cdot [\epsilon_f(\nabla\phi_a+\nabla\phi_d)-\epsilon_p(\nabla\phi_a+\nabla\bar \phi_d)],\\
	&=-\mathbf{r}/a\cdot [\\
	&\epsilon_f(\nabla\phi_a(\bzero)+\nabla\nabla\phi_a(\bzero)\cdot \br+\frac{1}{2}\nabla\nabla\nabla\phi_a(\bzero):\br\br\\
	&+\frac{\bP}{|\br|^3}-\frac{3\mathbf{r\cdot P}}{|\br|^5}\mathbf{r}+\frac{1}{2}(\frac{2\mathbf{Q\cdot r}}{|\br|^5}-\frac{5\mathbf{rr: Q}}{|r|^7}\mathbf{\br})\\
	&+\frac{1}{2}(\frac{3\bO:\br\br}{|\br|^7}-\frac{7\bO\tvdots\br\br\br}{|\br|^9}\mathbf{r}))\\
	&-\epsilon_p(\nabla\phi_a(\bzero)+\nabla\nabla\phi_a(\bzero)\cdot \br+\frac{1}{2}\nabla\nabla\nabla\phi_a(\bzero):\br\br\\
       &+\frac{\mathbf{ P}}{a^3}+\frac{\mathbf{Q\cdot r}}{a^5}+\frac{3}{2}\frac{\bO:\br\br}{a^7})],\\
	&=(\epsilon_p-\epsilon_f)\frac{\br\cdot \nabla\phi_a(\bzero)}{a}+(\epsilon_p-\epsilon_f)\frac{\br\br:\nabla\nabla\phi_a(\bzero)}{a}\\
	&+\frac{(\epsilon_p-\epsilon_f)}{2}\frac{\br\br\br\tvdots\nabla\nabla\nabla\phi_a(\bzero)}{a}\\
	&+(2\epsilon_f+\epsilon_p)\frac{\mathbf{r\cdot P}}{a^4}+(\frac{3}{2}\epsilon_f+\epsilon_p)\frac{\mathbf{rr: Q}}{a^6}\\
	&+(2\epsilon_f+\frac{3}{2}\epsilon_p)\frac{\br\br\br\tvdots \bO}{a^8}, \qquad |\mathbf{r}|=a.
\end{split}
\end{equation}
Here we have truncated 
at cubic terms in $|\br|$. 

Similarly, for the jump in the normal current 
\begin{equation}
\label{aj}
\begin{split}
	\bn\cdot[\bJ]&=-\mathbf{n}\cdot [\sigma_f(\nabla\phi_a+\nabla\phi_d)-\sigma_p(\nabla\phi_a+\nabla\bar \phi_d)],\\
	&=(\sigma_p-\sigma_f)\frac{\br\cdot \nabla\phi_a(\bzero)}{a}+(\sigma_p-\sigma_f)\frac{\br\br:\nabla\nabla\phi_a(\bzero)}{a}\\
		&+\frac{(\sigma_p-\sigma_f)}{2}\frac{\br\br\br\tvdots\nabla\nabla\nabla\phi_a(\bzero)}{a}\\
	&+(2\sigma_f+\sigma_p)\frac{\mathbf{r\cdot P}}{a^4}+(\frac{3}{2}\sigma_f+\sigma_p)\frac{\mathbf{rr: Q}}{a^6},\\
		&+(2\epsilon_f+\frac{3}{2}\epsilon_p)\frac{\br\br\br\tvdots \bO}{a^8}, \qquad |\mathbf{r}|=a.
\end{split}
\end{equation}
Substituting $\bu$ into the convection term, we have
\begin{equation}
\label{adsq}
\begin{split}
	\nabla_s\cdot (q \mathbf{u})&=q \nabla_s\cdot  (\mathbf{\Omega\times r})+(\mathbf{\Omega\times r})\cdot \nabla_s q \\
	&=(\mathbf{\Omega\times r})\cdot \nabla_s q.
\end{split}
\end{equation}
Then,
\begin{equation}
\label{adsq2}
\begin{split}
\nabla_s q&=\mathbf{(I-\bn\bn)\cdot \nabla}\\
&[(\epsilon_p-\epsilon_f)\frac{\br\cdot \nabla\phi_a(\bzero)}{a}+(\epsilon_p-\epsilon_f)\frac{\br\br:\nabla\nabla\phi_a(\bzero)}{a}\\
	&+\frac{(\epsilon_p-\epsilon_f)}{2}\frac{\br\br\br\tvdots\nabla\nabla\nabla\phi_a(\bzero)}{a}\\
	&+(2\epsilon_f+\epsilon_p)\frac{\mathbf{r\cdot P}}{a^4}+(\frac{3}{2}\epsilon_f+\epsilon_p)\frac{\mathbf{rr: Q}}{a^6}\\
	&+(2\epsilon_f+\frac{3}{2}\epsilon_p)\frac{\br\br\br\tvdots \bO}{a^8}]\\
&=\mathbf{(I-\hat{n}\hat{n})\cdot}\\
&[(\epsilon_p-\epsilon_f)\frac{\nabla\phi_a(\bzero)}{a}+2(\epsilon_p-\epsilon_f)\frac{\nabla\nabla\phi_a(\bzero)\cdot \br}{a}\\
&+\frac{3(\epsilon_p-\epsilon_f)}{2}\frac{\nabla\nabla\nabla\phi_a(\bzero):\br\br}{a}\\
&+(2\epsilon_f+\epsilon_p)\frac{\mathbf{ P}}{a^4}+(3\epsilon_f+2\epsilon_p)\frac{\bQ\cdot \br}{a^6}\\
&+(6\epsilon_f+\frac{9}{2}\epsilon_p)\frac{ \bO:\br\br}{a^8}].
\end{split}
\end{equation}
Thus
\begin{equation}
\label{adsq3}
\begin{split}
(\mathbf{\Omega\times r})\cdot \nabla_s q&=(\mathbf{\Omega\times r})\cdot \{\mathbf{(I-\bn\bn)\cdot }\\
&[(\epsilon_p-\epsilon_f)\frac{\nabla\phi_a(\bzero)}{a}+2(\epsilon_p-\epsilon_f)\frac{\nabla\nabla\phi_a(\bzero)\cdot \br}{a}\\
&+\frac{3(\epsilon_p-\epsilon_f)}{2}\frac{\nabla\nabla\nabla\phi_a(\bzero):\br\br}{a}\\
&+(2\epsilon_f+\epsilon_p)\frac{\mathbf{ P}}{a^4}+(3\epsilon_f+2\epsilon_p)\frac{\bQ\cdot \br}{a^6}\\
&+(6\epsilon_f+\frac{9}{2}\epsilon_p)\frac{ \bO:\br\br}{a^8}]\}
\end{split}
\end{equation}
\begin{equation}
\label{adsq4}
\begin{split}
=&-\mathbf{r}\cdot \{\mathbf{\Omega\times}[(\epsilon_p-\epsilon_f)\frac{\nabla\phi_a(\bzero)}{a}+(2\epsilon_f+\epsilon_p)\frac{\mathbf{ P}}{a^4}]\}\\
&-\mathbf{rr}:\{\mathbf{\Omega\times}[2(\epsilon_p-\epsilon_f)\frac{\nabla\nabla\phi_a(\bzero)}{a}+(3\epsilon_f+2\epsilon_p)\frac{\bQ}{a^6}]\}\\
&-\br\br\br\tvdots\{\mathbf{\Omega\times}[\frac{3(\epsilon_p-\epsilon_f)}{2}\frac{\nabla\nabla\nabla\phi_a(\bzero)}{a}+(6\epsilon_f+\frac{9}{2}\epsilon_p)\frac{ \bO}{a^8}]\}.
\end{split}
\end{equation}
Contributions from higher order moments can be added in the same way above.

Substituting back into \refeqq{qc}, the equation has contributions in different order of multipole products with $\br$. As we mentioned, $\br$ is a position vector at any point on the particle surface. The equation is separable at each order of $\br$ so we can obtain the independent equations for each multipole moment. When we only keep the dipole and quadrupole contribution in the system, we can obtain,

\begin{equation}
\label{ap1}
\begin{split}
&\frac{d}{d t}[(2\epsilon_f+\epsilon_p)\frac{\mathbf{r\cdot P}}{a^4}]+(2\sigma_f+\sigma_p)\frac{\mathbf{r\cdot P}}{a^4}+(\sigma_p-\sigma_f)\frac{\br\cdot \nabla\phi_a(0)}{a}\\
&- \mathbf{r}\cdot\{\mathbf{\Omega\times}[(\epsilon_p-\epsilon_f)\frac{\nabla\phi_a(0)}{a}+(2\epsilon_f+\epsilon_p)\frac{\mathbf{ P}}{a^4}]\}=0,
\end{split}
\end{equation}

\begin{equation}
\label{aq1}
\begin{split}
&\frac{d}{d t}[(\frac{3}{2}\epsilon_f+\epsilon_p)\frac{\mathbf{rr: Q}}{a^6}]+(\frac{3}{2}\sigma_f+\sigma_p)\frac{\mathbf{rr: Q}}{a^6}\\
&+(\sigma_p-\sigma_f)\frac{\br\br:\nabla\nabla\phi_a(0)}{a}-\mathbf{rr}:\{\mathbf{\Omega\times}[2(\epsilon_p-\epsilon_f)\frac{\nabla\nabla\phi_a(0)}{a}\\
&+(3\epsilon_f+2\epsilon_p)\frac{\bQ}{a^6}]\}=0.
\end{split}
\end{equation}

\refeqq{ap1} indicates

\begin{equation}
\label{ap2}
\frac{d \bP}{dt}=\mathbf{\Omega\times [P}+a^3\epsilon_{cm}\nabla\phi_{e}(0)]-\frac{1}{\tau_{mw}}[\bP+a^3\sigma_{cm}\nabla\phi_{e}(0)],
\end{equation}
where
\begin{equation}
\epsilon_{cm}=\frac{\epsilon_p-\epsilon_f}{\epsilon_p+2\epsilon_f}, \quad \sigma_{cm}=\frac{\sigma_p-\sigma_f}{\sigma_p+2\sigma_f},\quad \tau_{mw}=\frac{\epsilon_p+2\epsilon_f}{\sigma_p+2\sigma_f},\nonumber
\end{equation}

\begin{equation}
\label{aq2}
\begin{split}
\frac{d \mathbf{rr:Q}}{dt}=\mathbf{rr:\{\Omega \times } [2\bQ+4a^5\epsilon'_{cm}\nabla\nabla\phi_a(0)]\}\\
-\frac{1}{\tau'_{mw}}\mathbf{rr:[Q}+2a^5\sigma'_{cm}\nabla\nabla\phi_a(0)],
\end{split}
\end{equation}

where
\begin{equation}
\epsilon'_{cm}=\frac{\epsilon_p-\epsilon_f}{2\epsilon_p+3\epsilon_f}, \quad \sigma'_{cm}=\frac{\sigma_p-\sigma_f}{2\sigma_p+3\sigma_f}, \quad \tau'_{mw}=\frac{2\epsilon_p+3\epsilon_f}{2\sigma_p+3\sigma_f}.\nonumber
\end{equation}

\refeqq{aq2} indicates,
\begin{equation}
\label{aq4}
\begin{split}
\frac{d \mathbf{Q}}{dt}=&\mathbf{\{\Omega \times  [Q}+2a^5\epsilon'_{cm}\nabla\nabla\phi_a(0)]\}\\
&+\mathbf{\{\Omega \times  [Q}+2a^5\epsilon'_{cm}\nabla\nabla\phi_a(0)]\}^T\\
&-\frac{1}{\tau'_{mw}}\mathbf{[Q}+2a^5\sigma'_{cm}\nabla\nabla\phi_a(0)].
\end{split}
\end{equation}
Note we always assume $\bQ$ is symmetric to satisfy the original Laplace equation.

Also because
\begin{equation}
\label{atr1}
\mathbf{M=M^T}\quad \Rightarrow \quad tr(\mathbf{\Omega\times M})=0 \quad \forall\enskip \mathbf{\Omega,M\in \mathbfcal{R}_{3\times3}},
\end{equation}

\refeqq{aq4} guarantees $tr(\mathbf{Q})=0$. Additional details can be found in \cite{Hu2017}.

\section{Multi-particle system in linear electric fields}
In this section, we provide the details of the derivation of  equations of  motion of many particles.

The grand mobility formation for particle motion follows as,
	\begin{equation}
	\begin{split}
	\label{gm-1}
		\bordermatrix {
	&	\cr
    & \mathbf{u^{\infty}-u}       \cr
    & \mathbf{\Omega^{\infty}-\Omega}    \cr
}
=\mathbf{ M\enskip\cdot} 	\bordermatrix {
	&	\cr
    & \mathbf{\eta_f^{-1}\mathbf{F}}       \cr
    & \mathbf{\eta_f^{-1}\mathbf{T}}    \cr
}.
	\end{split}
	\end{equation}
	
For example of a two-particle system, the mobility equations of particle $1$ relates the hydrodynamic forces and torques to the particle motion as,
\begin{equation}
\begin{split}
\label{gm-2}
	&\bu^{\infty}-\bu_1=\eta_f^{-1}(\bf{a}_{11}\bF^H_1+\bf{a}_{12}\bF^H_2+\tilde{\bf{b}}_{11}\bT^H_1+\tilde{\bf{b}}_{12}\bT^H_2)\\
	&\bOmg^{\infty}-\bOmg_1=\eta_f^{-1}(\bf{b}_{11}\bF^H_1+\bf{b}_{12}\bF^H_2+\bf{c}_{11}\bT^H_1+\bf{c}_{12}\bT^H_2),
\end{split}
\end{equation}
where $\bT^H_i$ and $\bF^H_i$ are the torques and forces exerted by the fluid on particle i. $\eta_f$ is the viscosity of the external fluid. The coefficient tensors $\mathbf a$, $\bf b$, $\bf c$ are called mobility functions which are relative to particle separation.

Note that the full grand mobility matrix involves the strain if there is an applied shear flow. The model we present here is directly extendible by adding the applied strain contribution. The problem considered here have no applied shear flow. Such flows do occur in the multi-particle problem considered here but appear lower order in the analysis.

The coefficients in the mobility matrix are expanded as  \cite{Kim-Karrila:1991},
\begin{equation}
\begin{split}
(\mathbf{a}_{11})_{ij}&=x^a_{11}d_i d_j+y^a_{11}(\delta_{ij}-d_id_j),\\
(\mathbf{a}_{12})_{ij}&=x^a_{12}d_i d_j+y^a_{12}(\delta_{ij}-d_id_j),\\
(\mathbf{b}_{11})_{ij}&=y^b_{11}\epsilon_{ijk}d_k,\\
(\mathbf{b}_{12})_{ij}&=y^b_{12}\epsilon_{ijk}d_k,\\
(\mathbf{\tilde{b}}_{11})_{ij}&=(\mathbf{b}_{11})_{ji},\\
(\mathbf{\tilde{b}}_{12})_{ij}&=(\mathbf{b}_{21})_{ji}=-(\mathbf{b}_{12})_{ji},\\
(\mathbf{c}_{11})_{ij}&=x^c_{11}d_i d_j+y^c_{11}(\delta_{ij}-d_id_j),\\
(\mathbf{c}_{12})_{ij}&=x^c_{12}d_i d_j+y^c_{12}(\delta_{ij}-d_id_j),
\end{split}
\end{equation}
where $\mathbf{d}=\widehat{\bR}_{12}$. In a far-field approximation, assuming $\frac{a}{|\bR_{ij}|}\sim \gamma\ll1\enskip( \forall\enskip i,j)$, and denoting $R=|\bR_{ij}|$, the mobility functions are,
\begin{equation}
\begin{split}
	x^a_{11}&=\frac{1}{6\pi a}(1+\frac{15a^4}{4R^4}+O(\gamma^9)),\\
	y^a_{11}&=\frac{1}{6\pi a}(1+O(\gamma^6)),\\
	x^a_{12}&=\frac{1}{6\pi a}(\frac{3a}{2R}-\frac{a^3}{R^3}+O(\gamma^7)),\\
	y^a_{12}&=\frac{1}{6\pi a}(\frac{3a}{4R}+\frac{a^3}{2R^3}+O(\gamma^{11})),\\
	y^b_{11}&=O(\gamma^7),\\
	y^b_{12}&=\frac{1}{4\pi a^2}(-\frac{a^2}{2R^2}+O(\gamma^{10})),\\
	x^c_{11}&=\frac{1}{8\pi a^3}(1+O(\gamma^8)),\\
	x^c_{12}&=\frac{1}{8\pi a^3}(-\frac{a^3}{R^3}+O(\gamma^{11})),\\
	y^c_{11}&=\frac{1}{8\pi a^3}(1+O(\gamma^6)),\\
	y^c_{12}&=\frac{1}{8\pi a^3}(-\frac{a^3}{2R^3}+O(\gamma^9)),
\end{split}
\end{equation}
with the indicated leading order errors.

Then for multiple particles, the equations for particle motion will be($R_{ij}=|\bR_{ij}|$),
\begin{align}
\label{mobility}
	\eta_f(\bu^{\infty}-\bu_i)&=\frac{\bF^H_i}{6\pi a}+\sum_{j\neq i}[\frac{5a^3}{8\pi R^4_{ij}}(\bF^H_i\cdot\widehat{\bR}_{ij})\widehat{\bR}_{ij}\nonumber\\
	&-\frac{\bT^H_j\times \widehat{\bR}_{ij}}{8\pi R^2_{ij}}+\frac{1}{8\pi}(\frac{1}{R_{ij}}+\frac{2a^2}{3R^3_{ij}})\bF^H_j\nonumber\\
	&+\frac{1}{8\pi}(\frac{1}{R_{ij}}-\frac{2a^2}{R^3_{ij}})(\bF^H_j\cdot\widehat{\bR}_{ij})\widehat{\bR}_{ij}]\nonumber\\+O(\gamma^4),\\
\label{mobility2}
	\eta_f(\bOmg^{\infty}-\bOmg_i)&=\frac{\bT^H_i}{8\pi a^3}+\sum_{j\neq i}[-\frac{\bT^H_j}{16\pi R^3_{ij}}-\frac{\bF^H_j \times \widehat{\bR}_{ij}}{8\pi R^2_{ij}}]\nonumber\\
	&-\frac{3}{16\pi R^3_{ij}}(\bT^H_j\cdot\widehat{\bR}_{ij})\widehat{\bR}_{ij}+O(\gamma^6).	
\end{align}

Assume particles are forced balanced and no inertia effect is considered. The hydrodynamic force $\bF^H$ imposed by the fluid on particles should be balanced by the non-hydrodynamic interactions. i.e.
\begin{equation}
\label{fbalance}
	\mathbf{F}^H=-\bF^{el}-\bF^{rep}.
\end{equation}

For a given linear electric field, we can get calculate the force and torque exactly as,
\begin{equation}
\begin{split}
\label{ft0}
	&\bF_i^{el}=-4\pi\epsilon_f (\bP_i\cdot \nabla\nabla\phi_e(\mathbf{r_i})+\frac{1}{6}\bQ_i:\nabla\nabla\nabla\phi_e(\mathbf{r_i})),\\
&\bT_i^{el}=-4\pi\epsilon_f (\bP_i\times\nabla\phi_e(\mathbf{r_i})+(\bQ_i\cdot\nabla)\times\nabla\phi_e(\mathbf{r_i})),
\end{split}
\end{equation}
where $\phi_e=\phi_a+\sum\phi_d$ is the total external electric potential which contains the applied potential and the disturbance potentials.

Substitute the exact $\phi_e$ into \refeqq{ft0}, up to the order of $O(\gamma^4)$, we obtain three terms of the force:
\begin{equation}
\begin{split}
	\bF_i^{d1}=&4\pi\epsilon_f \bP_i\cdot \nabla\nabla \phi_a(\mathbf{r_i}),\\
	\bF_i^{d2}=&-\sum_{j\neq i}\frac{12\pi\epsilon_f}{R^4_{ij}}[(\bP_i\cdot \widehat{\bR}_{ij})\bP_j+(\bP_j\cdot\widehat{\bR}_{ij})\bP_i\\
	&+(\bP_i\cdot \bP_j)\widehat{\bR}_{ij}-5(\bP_j\cdot\widehat{\bR}_{ij})(\bP_i\cdot\widehat{\bR}_{ij})\widehat{\bR}_{ij}],\\
	\bF_i^{d3}=&\frac{2\pi\epsilon_f}{3} \bQ_i:\nabla\nabla\nabla\phi_a(\mathbf{r_i}).
\end{split}
\end{equation}
The leading error is from truncating the quadrupole contribution in the disturbance potential, hence
\begin{equation}
	\bF_i^{el}=\bF_i^{d1}+\bF_i^{d2}+\bF_i^{d3}+O(\gamma^5).
\end{equation}

Similarly when we deal with the hydrodynamic torque $T^H$, we assume it is instantly balanced by the electric torque, i.e.
\begin{equation}
	\mathbf{T}^H=-\bT^{el}.
\end{equation}
Then also from \refeqq{ft0},
\begin{equation}
	\bT_i^{el}=\bT_i^{d1}+\bT_i^{d2}+\bT_i^{d3}+\bT_i^{d4}+O(\gamma^5),
\end{equation}
where
\begin{equation}
\begin{split}
\label{multiT}
	\bT_i^{d1}=&4\pi\epsilon_f (\bP_i\times \nabla\phi_{a}(\mathbf{r_i})+(\bQ_i\cdot \nabla)\times\nabla\phi_{a}(\br_i)),\\
	\bT_i^{d2}=&-4\pi\epsilon_f \bP_i\times \sum_{j\neq i}(\nabla\frac{\br\cdot \bP_j}{r^3}\big |_{\br=\bR_{ij}}),\\
	\bT_i^{d3}=&-4\pi\epsilon_f \bP_i\times \sum_{j\neq i}(\nabla\frac{\br\br: \bQ_j}{2r^5}\big |_{\br=\bR_{ij}}),\\
	\bT_i^{d4}=&-4\pi\epsilon_f(\bQ_i\cdot \nabla)\times(\sum_{j\neq i}\nabla\frac{\br\cdot \bP_j}{r^3}\big |_{\br=\bR_{ij}}).
\end{split}
\end{equation}

Substitute \refeqq{fbalance}-\refeqq{multiT} into the mobility equation \refeqq{mobility}, we obtain the evolution equation for the particle translational velocity and angular velocity, 
\begin{equation}
\begin{split}
\label{motion}
	\bu_i&=\bu^{\infty}_i+\frac{\bF^{el}_i}{6\pi a\eta_f}+\sum_{j\neq i}\frac{\bF^{rep}_{ij}}{6\pi a\eta_f}+\sum_{j\neq i}\frac{(5\bF^{el}_i\cdot\widehat{\bR}_{ij})\widehat{\bR}_{ij}}{8\pi a\eta_f R^4_{ij}}\\
	&+\frac{1}{\eta_f}\sum_{j\neq i}[-\frac{\bT_j^{dep}\times \widehat{\bR}_{ij}}{8\pi R^2_{ij}}\\
	&+\frac{1}{8\pi}(\frac{1}{R_{ij}}+\frac{2a^2}{3R^3_{ij}})\bF^{el}_j\\
	&+\frac{1}{8\pi}(\frac{1}{R_{ij}}-\frac{2a^2}{R^3_{ij}})(\bF^{el}_j\cdot\widehat{\bR}_{ij})\widehat{\bR}_{ij}]+O(\gamma^5),\\
	\bOmg_i&=\bOmg^{\infty}_i+\frac{\bT_i^{el}}{8\pi a^3\eta_f}\\
	&+\frac{1}{\eta_f}\sum_{j\neq i}[-\frac{\bT_j^{dep}}{16\pi R^3_{ij}}-\frac{3}{16\pi R^3_{ij}}(\bT_j^{dep}\cdot\widehat{\bR}_{ij})\widehat{\bR}_{ij}\\
	&-\frac{(\bF^{el}_j) \times \widehat{\bR}_{ij}}{8\pi R^2_{ij}}]+O(\gamma^5).
\end{split}
\end{equation} 

As given above, the accuracy of both $\bu$ and $\bOmg$ is kept up to $O(\gamma^{4})$. The quadrupole contributes to both the DEP force and torque calculation. However, we need to clarify that while this calculation holds well for any linear electric field, for non linear fields it may not  be quite accurate. One reason was explained in the previous section that higher order moments are coupled into the equation when quadratic or higher order field components are non zero. The other reason is that for a rapidly or slowly changing field, the multipole moments have different magnitude scale. Then it is necessary to introduce another asymptotic parameter. In Appendix C, we will discuss a slowly varying electric field, which is more commonly seen in practical  applications.

\section{Slowly Varying Non-uniform Electric Fields}
For a single particle suspended in a general slowly varying electric field, we want to look at the asymptotic behavior when the particle radius is much smaller than the non-uniformity. The classic DEP force and torque calculation gives \refeqq{ft0} when the exact dipole and quadrupole moments are known. However, we would like to point out that, for a general electric field that induced non-zero octopole and higher moments, the error by truncating octopole moments comes at the same scale of the quadrupole contribution.

Suppose a potential $\phi_a$ is applied externally in a single particle suspension. The Taylor expansion of the applied electric field at the particle center is, 
\begin{equation}
\label{Etl1}
	\nabla \phi_a(\mathbf{r})=\nabla\phi_a(\bzero)+\mathbf{\nabla\nabla}\phi_a(\bzero)\cdot \br+\frac{1}{2}\nabla\nabla\nabla\phi_a(\bzero):\br\br+\cdots.
\end{equation} 

For slowly varying fields, we assume the length scale of the gradient operator is $L\gg a$, where $a$ is the particle radius. Thus we denote
\begin{equation}
-(\nabla)^{n}\phi_a\equiv \bE_a^{(n-1)}=\frac{E_c}{L^{n-1}}\widetilde\bE_a^{(n-1)},\quad n\geq 1,
\end{equation}
where $E_c$ is a characteristic electric field strength. i.e. $\widetilde\bE_a^{(0)}$ indicates the leading term of the scaled electric field and $\widetilde\bE_a^{(n)}$ is in general a tensor. All the $\widetilde\bE_a^{(n)}$ are dimensionless and $O(1)$.

In the following asymptotic analysis, we use the scaling scheme as,
\begin{equation}
\begin{aligned}
 &\tilde{t}=t/t_{ehd},\quad \widetilde{\mathbf{\Omega}}=\mathbf{\Omega} t_{ehd},\quad \tbr=\br/a,\\
 &\tbP=\frac{\bP}{E_c a^3},\quad \tbQ=\frac{\bQ}{E_c a^4},
\end{aligned}\nonumber
\end{equation}
where $t_{ehd}=\frac{\eta_f}{\epsilon_f E^2_c}$ is a characteristic EHD time scale.

Then the dimensionless form of the expansion \refeqq{Etl1} with the remainder term is,
\begin{equation}
\label{eadl}
\begin{split}
	\widetilde \bE_a(\tbr)=&\widetilde \bE_a^{(0)}(\bzero)+\delta\widetilde \bE_a^{(1)}(\bzero)\cdot \tbr+\frac{1}{2}\delta^2\widetilde\bE_a^{(2)}(\bzero): \tbr\tbr\\
	&+\frac{1}{6}\delta^3\widetilde\bE_a^{(3)}(\bzero)[\cdot]^{3} \tbr\tbr\tbr+\cdots,
\end{split}	
\end{equation} 
where $\delta=a/L\ll 1$ is a small asymptotic parameter.

Meanwhile, following \refeqq{pt3}, the dimensionless form of the induced field in the outer space has the expansion as,
\begin{equation}
\label{eddl}
\begin{split}
	\widetilde \bE_d(\tbr)&=-\frac{\widetilde \bP}{|\tbr|^3}+\frac{3\tbr\cdot \widetilde\bP}{|\tbr|^5}\tbr-\frac{\widetilde\bQ\cdot \tbr}{|\tbr|^5}+\frac{5\tbr\tbr: \widetilde\bQ}{2|\tbr|^7}\tbr\\
	&-\frac{3\widetilde\bO:\tbr\tbr}{2|\tbr|^7}+\frac{7\widetilde\bO\tvdots\tbr\tbr\tbr}{2|\tbr|^9}\tbr+\cdots.
\end{split}
\end{equation} 
From \refeqq{multiP}, when other particles are present, the induced potentials from other particles should be introduced into the total electric field. These disturbance fields contribute to the total external field as,
\begin{equation}
\label{esum}
\widetilde \bE_{e,i}=\widetilde \bE_{a,i}+\sum_{j\neq i}\widetilde \bE_{d,j}.
\end{equation}
Assuming the particles are widely separated, $\bE_{d,j}$ would be expanded in a far-field form. At the center of particle $i$, the field is
\begin{equation}
\tbE_{d,j}=-\frac{1}{|\tbR|^3}\Pi_1\widetilde \bP_j-\frac{1}{|\tbR|^4}\Pi_2\widetilde \bQ_j,
\end{equation}
where $\Pi_1\widetilde \bP_j=\widetilde \bP_j-3(\widetilde \bP_j\cdot \hbR)\hbR$ and $\Pi_2\widetilde \bQ_j=\widetilde \bQ_j\cdot\hbR-\frac{5}{2}(\widetilde \bQ_j: \hbR\hbR)\hbR$. Here we also have truncated the potential due to octopole and higher moments.

Now we encounter the second length scale, which is the particle separations $|\tbR|$. Assume a characteristic particle separation $\widetilde{R}_0\gg 1$. Denote $\gamma=1/\widetilde{R}_0$,  $\bbR=\tbR/\widetilde{R}_0$ and also assume all the particle separations are at the same scale, i.e. 
\begin{equation}
\frac{1}{|\tbR|}=\gamma\frac{1}{|\bbR|} \sim  O(\gamma). 
\end{equation}
Then we need to carefully select an appropriate asymptotic matching for the two small parameters $\delta$ and $\gamma$.

In order to incorporate particle interactions, we assume the balance as
\begin{equation}
\delta=\gamma^2,
\end{equation}
which indicates an even slower varying applied field than the particle disturbances.

Then we are able to expand the multipole moments in terms of the parameter $\gamma$, without causing fractal orders.
Still from \refeqq{multiP}, \refeqq{multiQ} and their derivation in the previous sections, we obtain,
\begin{equation}
\begin{split}
\label{exp}
&\widetilde \bP=\widetilde \bP^{(0)}+\gamma^{3}\widetilde \bP^{(3)}+\gamma^4\widetilde \bP^{(4)}+\cdots,\\
&\widetilde \bQ=\gamma^2\widetilde \bQ^{(2)}+\gamma^4\widetilde \bQ^{(4)}+\cdots,\\
&\widetilde \bO=\gamma^4\widetilde \bO^{(4)}+\cdots,\\
&\cdots\\
&\widetilde \bOmg=\widetilde \bOmg^{(0)}+\gamma^3\widetilde \bOmg^{(3)}+\gamma^4\widetilde \bOmg^{(4)}+\cdots.
\end{split}
\end{equation}

Assuming $\tbOmg^{\infty}=\bzero$, the rotation is actually determined by the multiple moments from \refeqq{motion}. In our balancing, the leading order nonzero contribution will be $\bOmg^{(0)}$ and the next nonzero orders should be $\bOmg^{(3)}$ and $\bOmg^{(4)}$.

Then each order of dipole moment satisfies the following evolution equations, 

$O(1)$:
 \begin{equation}
\label{PO1}
\frac{d \tbP^{(0)}}{dt}=\tbOmg^{(0)}\times [\tbP^{(0)}-\epsilon_{cm}\widetilde \bE_a^{(0)}(\bzero)]-\frac{1}{D}[\tbP^{(0)}-\sigma_{cm}\widetilde \bE_a^{(0)}(\bzero)];
\end{equation}

$O(\gamma^3)$:
 \begin{equation}
 \begin{split}
\label{PO3}
\frac{d \tbP^{(3)}}{dt}=&\tbOmg^{(0)}\times[\tbP^{(3)}+\epsilon_{cm}\frac{1}{|\bbR|^3}\Pi_1\widetilde \bP^{(0)}_j]+\tbOmg^{(3)}\times\tbP^{(0)}\\
&-\frac{1}{D}[\tbP^{(3)}+\sigma_{cm}\frac{1}{|\bbR|^3}\Pi_1\widetilde \bP^{(0)}_j];
\end{split}
\end{equation}

$O(\gamma^3)$:
 \begin{equation}
 \begin{split}
\label{PO4}
\frac{d \tbP^{(4)}}{dt}=&\tbOmg^{(4)}\times [\tbP^{(0)}-\epsilon_{cm}\widetilde \bE_a^{(0)}(\bzero)]+\tbOmg^{(4)}\times\tbP^{(0)}\\
&-\frac{1}{D}\tbP^{(4)}.
\end{split}
\end{equation}

The next correction to the dipole moment will be at $O(\gamma^4)$.
 
From \refeqq{aq4}, similarly we obtain, the leading nonzero quadrupole moment comes at the order of $O(\gamma^2)$, satisfying the equation below,

$O(\gamma^2)$:
\begin{equation}
\label{QO2}
\begin{split}
\frac{d \widetilde \bQ^{(2)}}{d\tilde{t}}=&\tbOmg^{(0)}\times [\widetilde \bQ^{(2)}-2\epsilon'_{cm}\widetilde \bE_a^{(1)}(\bzero)]\\
&+[\tbOmg^{(0)}\times [\widetilde \bQ^{(2)}-2\epsilon'_{cm}\widetilde \bE_a^{(1)}(\bzero)]]\tran\\
&-\frac{1}{D'}[\widetilde \bQ^{(2)}-2\sigma'_{cm}\widetilde \bE_a^{(1)}(\bzero)];
\end{split}
\end{equation}

$O(\gamma^4)$:
\begin{equation}
\label{QO4}
\begin{split}
\frac{d \widetilde \bQ^{(4)}}{d\tilde{t}}=&\tbOmg^{(0)}\times [\widetilde \bQ^{(4)}+2\epsilon'_{cm}\widetilde\nabla_{\bbR}(\frac{1}{|\bbR|^3}\Pi_1\widetilde \bP^{(0)}_j)]\\
&+[\tbOmg^{(0)}\times [\widetilde \bQ^{(4)}+2\epsilon'_{cm}\widetilde\nabla_{\bbR}(\frac{1}{|\bbR|^3}\Pi_1\widetilde \bP^{(0)}_j)]]\tran\\
&-\frac{1}{D'}[\widetilde \bQ^{(4)}+2\sigma'_{cm}\widetilde\nabla_{\bbR}(\frac{1}{|\bbR|^3}\Pi_1\widetilde \bP^{(0)}_j)].
\end{split}
\end{equation}
The next nonzero contribution is $\widetilde \bQ^{(6)}$ at the order of $O(\gamma^6)$.
 
Now continue to look at the expansions of the force and torque. The DEP force and torque are originally calculated directly by integrating the Maxwell stress tensor $\bf \Sigma$ over the particle surface,
\begin{equation}
\label{fandq}
\widetilde \bF=\frac{1}{4\pi}\iint_{|\tbr|=1} \mathbf{\widetilde\Sigma}\cdot\bn d\tilde S,\quad
\widetilde \bT=\frac{1}{4\pi}\iint_{|\tbr|=1} \tbr\times(\mathbf{\widetilde\Sigma}\cdot\bn) d\tilde S,
\end{equation}
where 
\begin{equation}
\mathbf{\widetilde\Sigma}=\widetilde \bE\widetilde \bE-\frac{1}{2}|\widetilde \bE|^2\bI,
\end{equation}
and the $\tbE$ is the total field,
\begin{equation}
\tbE=\tbE_e+\tbE_d,
\end{equation}
where $\tbE_e$ is the total external field vector from \refeqq{esum}.

For a spherical particle which has the standard multipole potential and exposed to a slowly varying external field, the force and torque equations are given in the exact forms as,
\begin{equation}
\begin{split}
\label{FTslow}
	&\tbF=\tbP\cdot \widetilde\nabla\tbE_e(\bzero)+\frac{1}{6}\tbQ:\widetilde\nabla\widetilde\nabla\tbE_e(\bzero)+\cdots,\\
&\tbT=\tbP\times\tbE_e(\bzero)+(\tbQ\cdot\widetilde\nabla)\times\tbE_e(\bzero)+\cdots,
\end{split}
\end{equation}

By substitution we obtain the force and torque in each order of $\gamma$.

\begin{equation}
\begin{split}
\label{Fasym}
\widetilde \bF_i=&\gamma^2\tbP^{(0)}_i\cdot\widetilde \bE^{(1)}_{a,i}(\bzero)+\gamma^5\tbP^{(3)}_i\cdot\widetilde \bE^{(1)}_{a,i}(\bzero)+\gamma^6\tbP^{(4)}_i\cdot\widetilde \bE^{(1)}_{a,i}(\bzero)\\
&+\frac{1}{6}\gamma^6\tbQ^{(2)}_i:\tbE_a^{(2)}(\bzero)\\
&-\gamma^4\sum_{j\neq i}\widetilde \bP^{(0)}_i\cdot\widetilde\nabla_{\bbR}(\frac{1}{|\bbR|^3}\Pi_1\widetilde \bP^{(0)}_j)\\
&+O(\gamma^{7}).
\end{split}
\end{equation}

\begin{equation}
\begin{split}
\label{Tasym}
\widetilde \bT_i=&\widetilde \bP^{(0)}_i\times\widetilde \bE^{(0)}_{a,i}(\bzero)+\gamma^3\tbP_i^{(3)}\times\tbE_{a,i}^{(0)}(\bzero)+\gamma^4\tbP_i^{(4)}\times\tbE_{a,i}^{(0)}(\bzero)\\
&+\gamma^4\tbQ_i^{(2)}\times\tbE_{a,i}^{(1)}(\bzero)\\
&-\gamma^3\sum_{j\neq i}\widetilde \bP^{(0)}_i\times\frac{1}{|\bbR|^3}\Pi_1\widetilde \bP^{(0)}_j\\
&+O(\gamma^5).
\end{split}
\end{equation}

The orders of error are kept at $O(\gamma^7)$ and $O(\gamma^5)$ respectively.

Adding up all the required terms in the order of accuracy, the evolution equations of $\tbP$ and $\tbQ$  are,
\begin{equation}
\label{asymP2}
\begin{split}
\frac{d \widetilde{\bP}_i}{d\tilde{t}}=&\widetilde{\mathbf{\Omega}}_i \times [\tbP_i-\epsilon_{cm}\widetilde \bE_a^{(0)}(\tbr_i)+\sum_{j\neq i}\frac{1}{|\tbR|^3}\Pi_1\widetilde \bP_j]\\
&-\frac{1}{D}[\tbP_i-\sigma_{cm}\widetilde \bE_a^{(0)}(\tbr_i)+\sum_{j\neq i}\frac{1}{|\tbR|^3}\Pi_1\widetilde \bP_j]+O(\gamma^5),
\end{split}
\end{equation}

\begin{equation}
\label{asymQ2}
\begin{split}
\frac{d \widetilde \bQ_i}{d\tilde{t}}=&\widetilde{\mathbf{\Omega}}_i\times [\widetilde \bQ_i-2\epsilon'_{cm}\delta\widetilde \bE_a^{(1)}(\tbr_i)]\\
&+[\tbOmg_i\times [\widetilde \bQ_i-2\epsilon'_{cm}\delta\widetilde \bE_a^{(1)}(\tbr_i)]]\tran\\
&-\frac{1}{D'}[\widetilde \bQ_i-2\sigma'_{cm}\delta\widetilde \bE_a^{(1)}(\tbr_i)]+O(\gamma^4).
\end{split}
\end{equation}

The rotation and velocity are determined from the grand-mobility matrix as well,

\begin{align}
\label{asymomg}
	\widetilde{\mathbf{\Omega}}_i&=\frac{\tbT_i}{8}-\sum_{j\neq i}[\frac{\tbF_j \times \hbR}{8 |\tbR|^2}\nonumber\\
	&+\frac{\tbT_j}{16 |\tbR|^3}+\frac{3}{16 |\tbR|^3}(\tbT_j\cdot\hbR)\hbR]+O(\gamma^5),
\end{align} 
	
\begin{align}
\label{asymu}
		\tilde \bu_i&=\tilde\bu^{\infty}_i+\frac{\tbF_i}{6}+\sum_{j\neq i}\frac{5}{8|\tbR|^4}(\tbF_i\cdot\hbR)\hbR\nonumber\\
		&-\sum_{j\neq i}[\frac{\tbT_j\times \hbR}{8 |\tbR|^2}+\frac{1}{8}(\frac{1}{|\tbR|}+\frac{2}{3|\tbR|^3})\tbF_j\nonumber\\
		&+\frac{1}{8}(\frac{1}{|\tbR|}-\frac{2}{|\tbR|^3})(\tbF_j\cdot\hbR)\hbR]+O(\gamma^7),
\end{align} 
where
\begin{align}
\label{Fterm}
	\tbF_i=&4 \delta\tbP_i\cdot \widetilde \bE_a^{(1)}(\tbr_i)\nonumber\\
	&-\sum_{j\neq i}\frac{12}{|\tbR|^4}[(\tbP_i\cdot \hbR)\tbP_j+(\tbP_j\cdot\hbR)\tbP_i\nonumber\\
	&+(\tbP_i\cdot \tbP_j)\hbR-5(\tbP_j\cdot\hbR)(\tbP_i\cdot\hbR)\hbR]\nonumber\\
		&+\frac{2}{3} \delta^2\tbQ_i :\widetilde \bE_a^{(2)}(\tbr_i),
\end{align}
\begin{align}
\label{Tterm}		
	\tbT_i=&4 \tbP_i\times\widetilde \bE_a^{(0)}(\tbr_i)\nonumber\\
	&-4 \tbP_i\times\sum_{j\neq i}(\frac{1}{|\tbR|^3}\Pi_1\cdot \tbP_j)\nonumber\\
	&+4\delta \tbQ_i\times\widetilde \bE_a^{(1)}(\tbr_i).
\end{align}
Thus in the case of a spatially-slowly varying field, equations \refeqq{asymP2}, \refeqq{asymQ2}, \refeqq{asymomg} and \refeqq{asymu} are the proper model to simulate  particle dynamics, where the error of the particle velocities is kept at $O(\widetilde{R}_0^{-7})$.\\

\end{document}